\documentclass[twocolumn]{aastex631}
\usepackage[utf8]{inputenc}

\usepackage[multidot]{grffile}

\usepackage[utf8]{inputenc}
\usepackage[T1]{fontenc}
\usepackage{ae,aecompl}

\DeclareRobustCommand{\VAN}[3]{#2}
\let\VANthebibliography\thebibliography
\def\thebibliography{\DeclareRobustCommand{\VAN}[3]{##3}\VANthebibliography}

\usepackage{graphicx}
\usepackage{amsmath}
\usepackage{amssymb}

\usepackage{comment}
\usepackage{color}
\usepackage{booktabs}
\usepackage{tabularx}
\usepackage{threeparttable}

\usepackage{academicons}
\usepackage{svg}
\newcommand{\orcid}[1]{\href{https://orcid.org/#1}{\textsuperscript{\includegraphics[width=10pt]{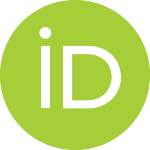}}}}

\newcommand{\mmode}[1]{\ifmmode{#1}\else{$#1$}\fi}

\newcommand{\Kepler}[0]{\emph{Kepler}}

\newcommand{\Gaia}[0]{\emph{Gaia}}
\newcommand{\Teff}[0]{\mmode{T_\text{eff}}}
\newcommand{\Rsolar}[0]{\mmode{\text{R}_{\odot}}}
\newcommand{\Msolar}[0]{\mmode{\text{M}_{\odot}}}
\newcommand{\Lsolar}[0]{\mmode{\text{L}_{\odot}}}

\newcommand{\logg}[0]{\mmode{\log g}}
\newcommand{\Dnu}[0]{\mmode{\Delta\nu}}

\newcommand{\numax}[0]{\mmode{\nu_\text{max}}}
\newcommand{\nuac}[0]{\mmode{\nu_\text{ac}}}
\newcommand{\DPi}[1]{\mmode{\Delta\Pi_{#1}}}

\newcommand{\muHz}[0]{\mmode{\mu{\rm Hz}}}

\newcommand{\fnumax}[0]{\mmode{f_{\nu_{\rm max}}}}
\newcommand{\dnum}[0]{\mmode{\delta\nu_{\rm m}}}

\newcommand{\Yinit}[0]{\mmode{Y_{\rm init}}}
\newcommand{\Xinit}[0]{\mmode{X_{\rm init}}}
\newcommand{\Zinit}[0]{\mmode{Z_{\rm init}}}
\newcommand{\amlt}[0]{\mmode{\alpha_{\rm MLT}}}
\newcommand{\fov}[1]{\mmode{f_{\rm ov, #1}}}
\newcommand{\chisq}[1]{\mmode{\chi^2_{#1}}}
\newcommand{\mh}[0]{\mmode{{\rm [M/H]}}}

\newcommand{\cyan}[1]{\textcolor{cyan} }

\usepackage{CJKutf8}
\newcommand{\CNnames}[1]{{\begin{CJK}{UTF8}{gbsn}~(#1)~\end{CJK}}}

\makeatletter
\newcommand\thefontsize[1]{{#1 The current font size is: \f@size pt\par}}
\makeatother

\defcitealias{liyg++2022-surface}{Li, Y. et~al.}
\defcitealias{liyg++2021-sc-intrinsic-scatter}{Li, Y. et~al.}
\defcitealias{litd++2022-kepler-rgb}{Li, T. et~al.}
\defcitealias{litd++2018-amlt}{Li, T. et~al.}
\defcitealias{litd++2020-kepler-36-subgiants}{Li, T. et~al.}
\newcommand{\mycitet}[1]{\citetalias{#1}~(\citeyear{#1})}
\newcommand{\mycitep}[1]{(\citetalias{#1}~\citeyear{#1})}
\newcommand{\mycitealt}[1]{\citetalias{#1}~\citeyear{#1}}

\newcommand{\rev}[1]{#1}

\begin{document}

\title{Realistic Uncertainties for Fundamental Properties of Asteroseismic Red Giants and the Interplay Between Mixing Length, Metallicity and \numax{}}

\author[0000-0003-3020-4437]{Yaguang Li\CNnames{李亚光}}
\affiliation{Institute for Astronomy, University of Hawai`i, 2680 Woodlawn Drive, Honolulu, HI 96822, USA}
\affiliation{Sydney Institute for Astronomy (SIfA), School of Physics, University of Sydney, NSW 2006, Australia}

\author[0000-0001-5222-4661]{Timothy R. Bedding}
\affiliation{Sydney Institute for Astronomy (SIfA), School of Physics, University of Sydney, NSW 2006, Australia}

\author[0000-0001-8832-4488]{Daniel Huber}
\affiliation{Institute for Astronomy, University of Hawai`i, 2680 Woodlawn Drive, Honolulu, HI 96822, USA}
\affiliation{Sydney Institute for Astronomy (SIfA), School of Physics, University of Sydney, NSW 2006, Australia}

\author[0000-0002-4879-3519]{Dennis Stello}
\affiliation{School of Physics, University of New South Wales, 2052, Australia}
\affiliation{Sydney Institute for Astronomy (SIfA), School of Physics, University of Sydney, NSW 2006, Australia}

\author[0000-0002-4284-8638]{Jennifer van Saders}
\affiliation{Institute for Astronomy, University of Hawai`i, 2680 Woodlawn Drive, Honolulu, HI 96822, USA}

\author[0000-0003-0817-6126]{Yixiao Zhou\CNnames{周一啸}}
\affiliation{Stellar Astrophysics Centre, Department of Physics and Astronomy, Aarhus University, Ny Munkegade 120, DK-8000 Aarhus C, Denmark}

\author[0000-0002-7654-7438]{Courtney L. Crawford} %
\affiliation{Sydney Institute for Astronomy (SIfA), School of Physics, University of Sydney, NSW 2006, Australia}

\author[0000-0002-8717-127X]{Meridith Joyce}
\affiliation{Konkoly Observatory, HUN-REN Research Centre for Astronomy and Earth Sciences, Konkoly-Thege Mikl\'os \'ut 15-17, H-1121, Budapest, Hungary}
\affiliation{CSFK, MTA Centre of Excellence, Budapest, Konkoly-Thege Mikl\'os \'ut 15-17, H-1121, Budapest, Hungary}

\author[0000-0001-6396-2563]{Tanda Li\CNnames{李坦达}}
\affiliation{Department of Astronomy, Beijing Normal University, Haidian District, Beijng 100875, China}

\author[0000-0002-5648-3107]{Simon J. Murphy}
\affiliation{Centre for Astrophysics, University of Southern Queensland, Toowoomba, QLD 4350, Australia}

\author[0000-0003-1179-2069]{K. R. Sreenivas}
\affiliation{Sydney Institute for Astronomy (SIfA), School of Physics, University of Sydney, NSW 2006, Australia}


\begin{abstract}
Asteroseismic modelling is a powerful way to derive stellar properties. However, the derived quantities are limited by built-in assumptions used in stellar models. 
This work presents a detailed characterisation of stellar model uncertainties in asteroseismic red giants, focusing on the mixing-length parameter \amlt{}, the initial helium fraction \Yinit{}, the solar abundance scale, and the overshoot parameters. 
First, we estimate error floors due to model uncertainties to be $\approx$0.4\% in mass, $\approx$0.2\% in radius, and $\approx$17\% in age, primarily due to the uncertain state of \amlt{} and \Yinit{}. 
The systematic uncertainties in age exceed typical statistical uncertainties, suggesting the importance of their evaluation in asteroseismic applications. 
Second, we demonstrate that the uncertainties from \amlt{} can be entirely mitigated by direct radius measurements or partially through \numax{}. Utilizing radii from \Kepler{} eclipsing binaries, we determined the \amlt{} values and calibrated the \amlt{}--[M/H] relation. The correlation observed between the two variables is positive, consistent with previous studies using 1-D stellar models, but in contrast with outcomes from 3-D simulations.
Third, we explore the implications of using asteroseismic modelling to test the \numax{} scaling relation. We found that a perceived dependency of \numax{} on [M/H] from individual frequency modelling can be largely removed by incorporating the calibrated \amlt{}--[M/H] relation. Variations in \Yinit{} can also affect \numax{} predictions. These findings suggest that \numax{} conveys information not fully captured by individual frequencies, and that it should be carefully considered as an important observable for asteroseismic modelling.
\end{abstract}


\section{Introduction}\label{sec:intro}

Asteroseismolgy is the study of stellar structure via stellar oscillations.
Ground-based radial velocity instruments and space missions such as CoRoT, Kepler, and TESS have provided extensive data on millions of solar-type stars and red giants.
These stars show rich spectra of oscillation modes, enabling a detailed charaterisation of stellar interiors \citep{chaplin+2013-solar-like-review,basu+2017-book}.

Driven by advances in observational data, asteroseismology has rapidly become an essential tool for determining the fundamental properties of field stars, such as age, which are difficult to assess with other methods. 
One widely used approach, known as forward modelling, involves computing stellar structural models and their standing wave frequencies under varying initial conditions (like mass and chemical composition) and evolving them over time. These models are then matched with asteroseismic parameters derived from observational data.

In red giants, two main types of oscillation modes are excited: pressure (p) modes, which are sound waves probing the envelope cavity, and gravity (g) modes, which are buoyancy waves probing the core cavity. 
Asymptotic theory is useful for categorizing these oscillation frequencies \citep{tassoul-1980-asymptotic-relation,gough-1986-aymptotic-relation}. 
For p modes with radial order $n_p$ and spherical degree $l$, the frequencies are approximately evenly spaced: 
\begin{equation}
    \nu_{n,l} \approx \Delta\nu \left( n_p + \frac{l}{2} + \epsilon_p \right) + \delta\nu_{0,l}.
\end{equation}
Here, \Dnu{} is the p-mode large separation, which relates to the stellar mean density, $\delta\nu_{0,l}$ is the small separation between modes of different $l$-degrees, and $\epsilon_p$ is a phase offset.
For g modes with radial order $n_g$ and spherical degree $l$, the periods are also roughly evenly spaced:
\begin{equation}
    \frac{1}{\nu_{n,l}} \approx \Delta\Pi_l \left( n_g + \epsilon_{g,l}\right).
\end{equation}
Here, \DPi{l} is the g-mode period spacing, and $\epsilon_{g,l}$ is a phase offset. 
Because these oscillation modes directly probe their respective oscillation cavities, their frequencies or asymptotic parameters are sensitive to the evolving internal structure of stars, providing constraints for stellar ages \citep{davies++2016-age-review}.

In main-sequence dwarfs, the small separation $\delta\nu_{02}$ is sensitive to the chemical gradient near the core, which changes due to hydrogen burning \citep{jcd-1984-review}. 
In subgiants, the core contracts while the envelope expands, leading to a phenomenon where g modes in the core couple with p modes in the envelope, with a feature known as an avoided crossing \citep{deheuvels++2010-avoided-crossing,ong++2020-coupling}. 
At this stage, both the g mode frequencies and the coupling strength change rapidly, making this feature a valuable indicator of stellar age \mycitep{litd++2020-kepler-36-subgiants}.
In red giants, age largely depends on the duration of the main-sequence phase, which is strongly correlated with stellar mass. Since the sizes of the core and envelope are heavily influenced by the stellar mass and radius, both p and g modes are effective tools for estimating the ages of red giants.

Moreover, the observational data also allow for the extraction of amplitudes and damping rates of oscillation modes. However, current models face difficulties in predicting these values due to the complex interaction between convection and pulsation. As a result, the frequency of maximum power, \numax{}, is primarily used to constrain models. This is achieved through the application of a scaling relation: $\numax{}\propto g/\sqrt{\Teff{}}\propto MR^{-2}T_{\rm eff}^{-1/2}$, where $g$, $M$, $R$, and \Teff{} are the surface gravity, mass, radius, and effective temperature, respectively \citep{brown++1991-dection-procyon-scaling-relation,kjeldsen+1995-scaling-relations}. 
Despite its extensive application, this scaling relation remains empirical, and its limitations and broader applicability are not fully understood. These issues will be further explored in Section~\ref{sec:numax}.

Using these oscillation frequencies (or asymptotic parameters) and \numax{} as constraints, previous studies have consistently reported typical uncertainties of $\approx$4\% in mass, $\approx$2\% in radius, and $\approx$10\% in age for solar-type main-sequence stars and subgiants (e.g. \citealt{silvaaguirre++2015-33-kepler-exoplanet-host,silvaaguirre++2017-legacy2-modelling}; \mycitealt{litd++2020-kepler-36-subgiants}). 
Similar levels of uncertainty, particularly precision, have also been indicated for red giants (\citealt{montalban++2021-age}; \mycitealt{litd++2022-kepler-rgb}).
This marks a significant advancement over traditional isochrone fitting methods, which typically yield uncertainties of $\approx$10\% in mass and $\approx$50\% in age \citep{tayar++2022-error}.

Given the promising results of asterseismology, it is crucial to carefully examine the inherent systematic uncertainties in the underlying stellar models. 
Such uncertainties include the mixing-length parameter \amlt{} and initial helium abundance \Yinit{}, both essential for building stellar models. \amlt{} is a key parameter in modelling convection, which governs energy transport; \Yinit{} shapes the overall chemical composition, affecting the equation of state, opacity, and nuclear reaction rates. 
However, these parameters can not be determined unambiguously for field stars through observational methods \citep[e.g.][]{lebreton+2014-hd52265,appouchaux++2015-hip93511-binary,silvaaguirre++2015-33-kepler-exoplanet-host}.
One common approach adopted by popular stellar isochrones is to use solar-calibrated values \citep{demarque++2004-yy,dotter++2008-dsep,serenelli++2013-garstec,Choi++2016-mist-1-solar-scaled-models,nguyen++2022-parsec2}, which may not be suitable for the study of field stars given the availability of high-precision observational data, \rev{such as those on the lower main sequence \citep{cheny++2014-parsec} and the red-giant branch \citep{tayar-2017-amlt,joyce+2018-amlt-metal-poor}}. 
An alternative method, often employed in the asteroseismic community, involves treating these variables as free parameters in model grid construction, but this practice can obscure other model flaws and still lead to biases in the estimation of stellar properties \citep{cunha++2021-modelling-errors}. 
In practice, the bias in estimating stellar age becomes particularly important in the context of Galactic archaeology, where age scales for millions of red giants have been calibrated based on asteroseismic datasets \citep{ho++2017-lamost,wuyq++2018-mass-rgb,wangc++2023-lamost-age,anders++2023-apogee-age}.

In this paper, we aim to scrutinize some of the systematic uncertainties in models, and highlight their significant contribution to the overall uncertainty in determining fundamental stellar properties, especially age, which often surpasses statistical uncertainties. 
We will assess their impact on estimating stellar properties (Section~\ref{sec:property}), investigate the interplay between \amlt{}, radius and metallicity [M/H] (Section~\ref{sec:amlt}), and finally, discuss how these model uncertainties affect the validation of the \numax{} scaling relation (Section~\ref{sec:numax}).

\section{Data analysis}\label{sec:analysis}

\begin{figure}
    \centering
    \includegraphics{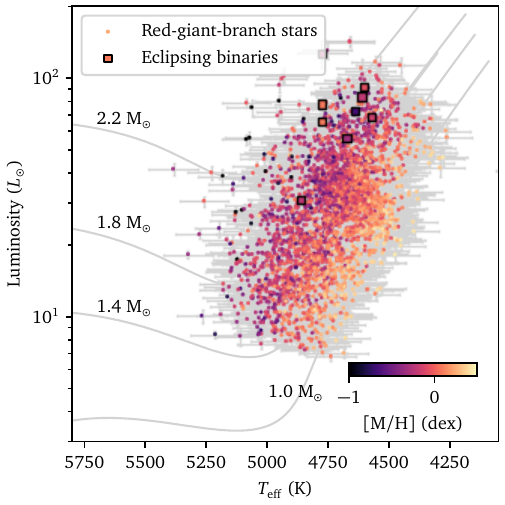}
    \caption{H--R diagram showing the sample used in this work, colour-coded by metallicity. A few solar-metallicity tracks are also shown for visual guidance. }
    \label{fig:hrd}
\end{figure}

\subsection{Observations}

We first describe the observational data used to constrain stellar models. In this work, we chose the low-luminosity asteroseismic red giants observed by \Kepler{} \citep{yuj++2018-16000-rg}. 
We selected our sample (below about 100~\Lsolar) by restricting \numax{} to be above 17~\muHz{}.
These low-luminosity red giants are on the red-giant branch (RGB), so contamination from asymptotic-giant-branch (AGB) stars is not a concern. Compared to high-luminosity red giants, these stars exhibit many orders of excited modes, with some of the most comprehensive asteroseismic constraints among all stellar types \citep{mosser++2011-uni-pattern}. 

For asteroseismic inputs, we collected global oscillation parameters, including \numax{}, \Dnu{} and \DPi{1}, which are commonly used for optimising stellar models when individual frequencies are unavailable. 
We chose \Dnu{} values determined by the SYD pipeline \citep{yuj++2018-16000-rg}, \numax{} values processed by the nuSYD pipeline \citep{sreenivas++2024-numax}. The nuSYD pipeline offers slightly improved precision in \numax{} compared to the SYD pipeline \citep{huber++2009-syd-pipeline,yuj++2018-16000-rg,chontos++2021-pysyd}, due to their simple treatment of granulation background and the low-level noise in their adopted PDCSAP (Pre-search Data Conditioning Simple Aperture Photometry) light curves \citep{stumpe++2012-pdc-1,smith++2012-pdc-2}. 
We used $l=1$ g-mode period spacings (\DPi{1}) from \citet{vrard++2016-period-spacings}, rather than the individual $l=1$ oscillation frequencies, to avoid the complexity of treating mixed modes in stellar modelling \citep{ball++2018-surface-effect-rgb}.

For the asteroseismic inputs, we also gathered individual frequencies, which are essential for fully utilizing all frequency information from oscillation spectra. We used oscillation frequencies ($l=0,2$) extracted by \citet{kallinger++2019-peakbagging}, who followed a standard peakbagging approach. 
However, we did not use the frequencies for $\ell=1$ modes, because they are mixed modes, and would require significant computational resources, especially for properly addressing the surface effect \citep{ong++2021-surf-corr-theory,ong++2021-surf-corr-sg}.

For spectroscopic inputs, we used metallicity [M/H] from APOGEE \citep{pinsonneault++2018-apokasc,abdurrouf++2021-apogee-dr17}, and \Teff{} that we derived from the IRFM method \citep{casagrande++2021-irfm-gaia}, which uses \Gaia{} and 2MASS photometry and extinction values from \citet{green++2019-dustmap} as inputs. 
By comparing the IRFM \Teff{} and the APOGEE \Teff{} values, we observed a systematic variation: the IRFM \Teff{} is 50~K higher than the APOGEE values at $\mh{}\sim-0.6$~dex, and conversely, 50~K lower at $\mh{}\sim0.4$~dex. The root-mean-square difference between the two \Teff{} sources is about 70~K across all [M/H] levels.
The differences in \Teff{} scales are likely related to how they were calibrated, rather than being physical. 
We found that the choice of \Teff{} scale made little difference to our results (see Section~\ref{sec:amlt}).

We determined luminosities using \emph{K$_s$}-band magnitudes \citep{cutri++2003-2mass-ps}, along with \citet{green++2019-dustmap} extinctions, \citet{Choi++2016-mist-1-solar-scaled-models} bolometric corrections, and \Gaia{} DR3 distances \citep{bailer-jones++2021-edr3-distances}. 

Furthermore, we included oscillating eclipsing binaries analysed by \citet{gaulme++2016-eb-sc}, \citet{brogaard++2018-accuracy-scaling-relation}, and \citet{benbakoura++2021-binary}, which have masses and radii measured from dynamical modelling. 
In cases where a star was analysed by multiple studies, we opted for the parameters reported from the most recent study. 
Figure~\ref{fig:hrd} presents the sample used in this work on the Hertzsprung-Russell diagram.

\subsection{Stellar models}

We constructed a new set of stellar models with MESA \citep[version r15140;][]{paxton++2011-mesa,paxton++2013-mesa,paxton++2015-mesa,paxton++2018-mesa,paxton++2019-mesa} and GRYE \citep[version 6.0.1;][]{townsend+2013-gyre}. The construction of these models is largely based on the input physics outlined in \mycitet{liyg++2022-surface}. The convection was formulated with the mixing-length theory from \citet{henyey++1965-mlt} and the surface boundary conditions were constructed with Eddington $T-\tau$ grey atmosphere model \citep{eddington-1926-star}. At the end of the paper, we provided a link to the configuration files used for producing the MESA models.

There are two primary differences compared to the \mycitet{liyg++2022-surface} models. Firstly, the new models considered possible variations in the convective overshoot. We used the exponential overshooting scheme according to \citet{herwig-2000-overshoot}, and varied the amount of overshoot for core and shell convective boundaries independently as $\fov{core}\in(0., 0.03)$ and $\fov{shell}\in(0., 0.02)$, respectively. The other free parameters for the grid were initial stellar mass $M\in(0.6, 2.5)$, metallicity $\mh\in(-1.0, 0.6)$, the initial helium abundance $\Yinit{}\in(0.20, 0.37)$, and the mixing-length parameter $\amlt{}\in(1.0, 2.7)$. No mass loss was included in the models.

Secondly, in addition to calculating the frequencies of the radial modes ($\ell=0$) we also calculated frequencies of decoupled (pure p) quadrupolar modes ($\ell=2$) using the method introduced by \citet{ong++2020-coupling}. We incorporated the $\ell=2$ modes in this work because the spacing between $\ell=0$ and $\ell=2$ modes, the so-called small separation, relates closely to stellar mass in red giants and could offer extra constraints on stellar properties \citep{montalban++2010-models,kallinger++2012-epsp}.

\subsection{Model fitting}

We followed the model-fitting framework described in \mycitet{liyg++2022-surface}. Each model is associated with global stellar parameters $\{M, R, L, {\rm Age}, \DPi{1}, \Teff{}, {\rm [M/H]}, ...\}$ and oscillation frequencies $\{\nu_i\}$. Following \mycitet{liyg++2022-surface}, we corrected the surface effect in $\{\nu_i\}$ with an ensemble approach, which helps eliminate unrealistic surface corrections and reduces scatter in model-derived properties. We achieved this by parameterizing the amount of surface correction at \numax{}, \dnum{}, and at 10\% above \numax{}, $\delta\nu_m'$, as functions of stellar surface properties:
\begin{equation}
    \dnum = a \cdot (g/{g_\odot})^{b} \cdot (T_{\rm eff}/T_{\rm eff,\odot})^{c} \cdot (d \cdot {\rm [M/H]} + 1),
\end{equation}
and
\begin{equation} 
\label{eq:dnumf}
\delta\nu_{m}' = a' \cdot (g/{g_\odot})^{b'} \cdot (T_{\rm eff}/T_{\rm eff,\odot})^{c'} \cdot (d' \cdot {\rm [M/H]} + 1).
\end{equation}
These two equations were then used to determine the surface terms ($a_{-1}$ and $a_{3}$) in the inverse-cubic correction formula \citep{ball+2014-surface-correction-inertia-weighted}:
\begin{equation}
    \delta\nu(\nu; a_{-1}, a_{3}) = \left(a_{-1}\nu^{-1} + a_{3}\nu^3 \right) / \mathcal{I},
\end{equation}
where $\mathcal{I}$ is the mode inertia. The parameters appearing in these equations, $\{a,b,c,d,a',b',c',d'\}$, were jointly fitted to the entire sample, yielding best-fitting values of $\{-6.11\pm0.28,0.79\pm0.04,,-5.04\pm0.85,-0.79\pm0.09,-7.69\pm0.72,0.79\pm0.03,-4.59\pm0.68,-0.87\pm0.08\}$. These values are slightly different from those reported by \mycitet{liyg++2022-surface}, due to the differences in underlying models and observational constraints, indicating the importance of such re-calibration for this method.

To determine stellar properties, we applied a range of observational constraints, quantified using chi-squared (\chisq{}) functions for goodness of fit:
\begin{equation}
    \chisq{q} = \left( \frac{q_{\rm obs}-q_{\rm mod}}{\sigma_{q}} \right)^2,
\end{equation}
where $q$ represents the observables \numax{}, \DPi{1}, $L$, \Teff{}, \mh{}, $M$, and $R$. 
For individual frequencies, we used reduced \chisq{} functions for a group of modes with the same $\ell$-degree, which are the standard \chisq{} functions averaged by the number of oscillation modes:
\begin{equation}
    \chisq{l} = \frac{1}{N_{l}} \sum_i^{N_{l}} \left( \frac{\nu_{{\rm obs}, l, i} - \nu_{{\rm mod}, l, i}}{\sigma_{\nu_{\rm obs}, l, i}} \right)^2.
\end{equation}
In a strict statistical sense, using an average by the number of modes is not appropriate, because each frequency was determined independently. However, these \chisq{} functions are very sensitive to inaccurate predictions of mode frequencies, due to the relatively small magnitudes of $\sigma_{\nu_{\rm obs}}$. Relying on a single set of grid models with a specific selection of input physics (as was done in this work) is unable to fully capture all sources of systematic uncertainties. This limitation could result in extremely large values for the seismic \chisq{} functions. Hence, the average method reduces the reliance of the seismic \chisq{} on the inaccurate predictions, and quantitatively it is similar to adding an extra error term for systematic uncertainties. We refer the reader to \citet{cunha++2021-modelling-errors} for a thorough discussion on this topic.

For illustration, we combine observational constraints into separate groups. The \chisq{\rm freq} includes individual frequencies for modes of $l=0$ and $l=2$:
\begin{equation}
    \chi^2_{\rm freq} = \chisq{l=0} + \chisq{l=2}.
\end{equation}
The seismic constraint \chisq{\rm seis} incorporates both individual frequencies and global seismic parameters (\numax{} and \DPi{1}):
\begin{equation}
    \chi^2_{\rm seis} = \chisq{\rm freq} + \chisq{\numax} + \chisq{\DPi{1}}.
\end{equation}
The \Gaia{} constraint applies to luminosity:
\begin{equation}
    \chisq{\rm Gaia} = \chisq{L},
\end{equation}
and the spectroscopic constraint covers \Teff{} and \mh{}:
\begin{equation}
    \chisq{\rm spec} = \chisq{\Teff} + \chisq{\mh}.
\end{equation}
For the eclipsing binaries, the dynamical constraint includes mass and radius obtained from the modelling of the binary orbit:
\begin{equation}
    \chi^2_{\rm dyn} = \chisq{M} + \chisq{R}.
\end{equation}
When combining these individual constraints, we implicitly assume that each was derived independently, even though this may not be the case. For example, \Teff{} and \mh{} are often strongly correlated from spectroscopic analysis. To accurately account for such correlations, it is essential to include information on the covariance matrix \citep[e.g.][]{gent++2022-sapp}. However, this information was not available for most of our sample.

Under the Bayesian model-fitting framework, the posterior probability for the model parameters is expressed as
\begin{equation}\label{eq:posterior}
    p(\theta | \mathcal{D}) = p(\theta) \times \mathcal{L}(\mathcal{D} | \theta),
\end{equation}
where $p(\theta)$ is the prior on model parameters, which are uniform within the grid boundaries unless noted otherwise.
The likelihood function quantifies the agreement between models (specified by $\theta$) and observational data ($\mathcal D$):
\begin{equation}
    \mathcal{L}(\mathcal{D} | \theta) = \exp\left(-\chi^2/2 \right),
\end{equation}
where \chisq{} includes various observational constraints that are detailed in subsequent sections.
The estimation of a stellar parameter of interest was determined by integrating Eq.~\ref{eq:posterior} over other model parameters, a process called marginalisation.

\section{Uncertainties from model input physics}\label{sec:property}

\begin{figure}
    \centering
    \includegraphics{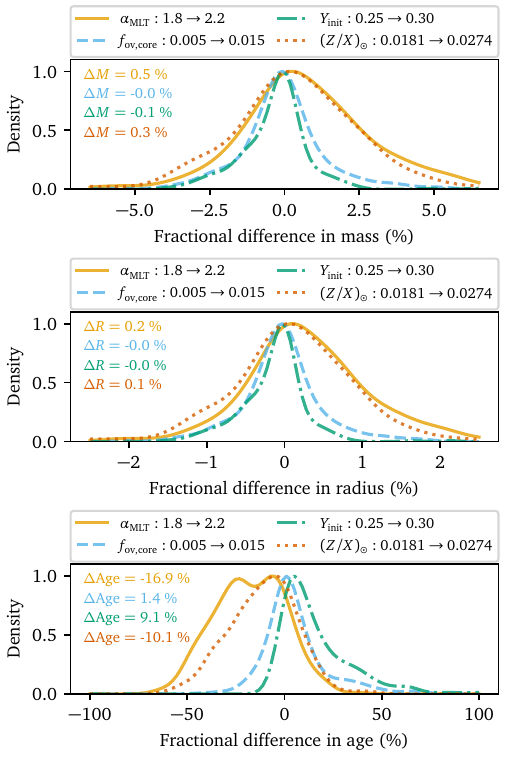}
    \caption{Distributions of the fractional differences in masses, radii, and ages for our red giant sample that result from changing the input physics. Each line corresponds to changing the values of a different model parameter (see legends), and the numbers indicate to the median differences. The distributions have been processed through Gaussian kernel density estimations for clearer visualization. The stellar properties shown in this diagram were derived with $\nu_{\ell=0}$, $\nu_{\ell=2}$, \DPi{1}, \numax{}, \Teff{}, \mh{}, and $L$.}
    \label{fig:physics}
\end{figure}

\begin{figure}
    \centering
    \includegraphics{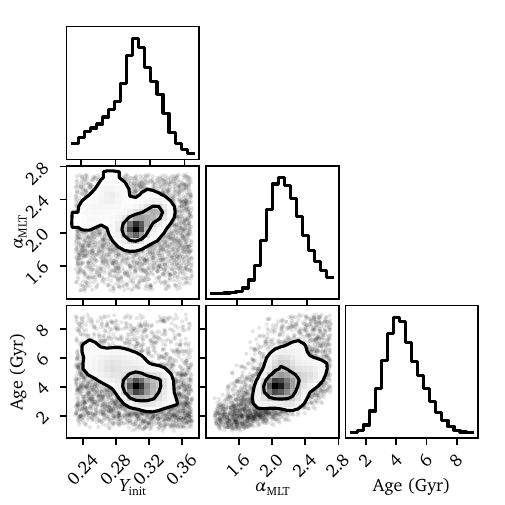}
    \caption{\rev{Posterior distributions of stellar models constrained by the observational properties of KIC 10904520. Each point represents a stellar model, colour-coded according to its probability. 1- and 2-$\sigma$ volumes are displayed as contours. } }
    \label{fig:corner}
\end{figure}

\begin{figure}
    \centering
    \includegraphics{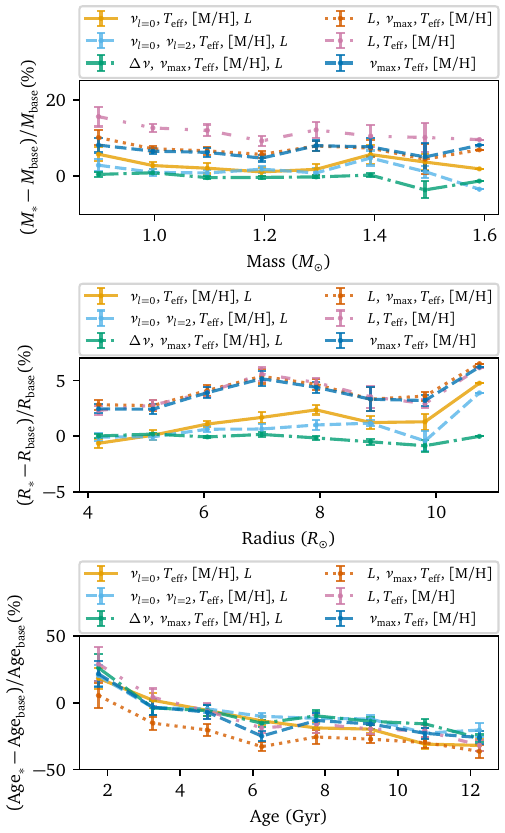}
    \caption{Changes in the derived masses, radii, and ages for our red giant sample that result from changing the input physics. Each point shows the binned medians and the associated errors on the binned medians. In each panel, the x-axis shows the stellar properties derived with $\nu_{\ell=0}$, $\nu_{\ell=2}$, \DPi{1}, \numax{}, \Teff{}, \mh{}, and $L$, denoted as the ``base'' case. They are compared against the stellar properties derived by other combinations of observational constraints (see legends). }
    \label{fig:observables}
\end{figure}

\subsection{Varying input physics}
When studying the effects of varying input physics, \citet{lebreton++2014-age} identified the most important factors contributing to the uncertainty in the main-sequence lifetime.
These include chemical abundances, convective core overshoot, and rotation, with discrepancies surpassing 30\% in comparison to a reference model. 
Chemical abundances and convective core overshoot are especially crucial for low-mass stars \rev{above 1.2~\Msolar{}} \citep[see also][]{ying++2023-m92,joyce++2023-age}.
In this section, we look deep into how variations in these two types of input physics influence the inferred stellar properties, in the context of using constraints from asteroseismology.

This investigation was done by adjusting key parameters in our models: the mixing-length parameter \amlt{}, the initial helium fraction \Yinit{}, the solar abundance values $(Z/X)_\odot$ used in metallicity calculations, and the core overshoot parameter, \fov{core}. 
Each star was fitted with the full set of observables: $\chisq{} = \chisq{\rm seis} + \chisq{\rm Gaia} + \chisq{\rm spec}$.
We assessed the fractional differences in stellar properties resulting from these adjustments, with the distributions displayed in Figure~\ref{fig:physics}.

\subsubsection{Mixing length}

A key focus was on the impact of varying \amlt{}, which prescribes convective flux transport in convection zones and is especially important in super-adiabatic regions. 
Although solar-calibrated \amlt{} values are commonly used in stellar isochrones, several studies advocate for and validate non-solar values across diverse samples \citep[see][]{joyce+2023-mlt}. 
In our analysis, we varied \amlt{} from 1.8 to 2.2, which covers the range commonly explored in these studies for red giants. 
The adjustment was implemented by modifying the priors in Eq.~\ref{eq:posterior}.
For example, setting \amlt{} to 1.8 involves applying the following normal distribution:
\begin{equation}
    \amlt \sim \mathcal{N} ( 1.8, \sigma),
\end{equation}
where $\mathcal{N}(\mu, \sigma)$ denotes a normal distribution with mean $\mu$ and standard deviation $\sigma$. Here we chose $\sigma$ as 0.05, and tests showed that using other values (0.025 and 0.1) did not lead to significant differences. 

As can be seen from Figure~\ref{fig:physics}, altering \amlt{} by 0.4 induces the largest variations among the tests we conducted. The median differences are 0.4\% in mass, 0.2\% in radius, and 16.4\% in age. The pronounced \amlt{} effect on age is indirect, because an increased \amlt{} tends to favour models with different abundance values that modify ages directly. Specifically, the 0.4 increase in \amlt{} decreases \Yinit{} by $\approx$0.08, leading to a reduced hydrogen burning rate on the main sequence \citep{lebreton++2014-age,mckkever++2019-ngc6791}. Moreover, since the metal-to-hydrogen ratio ([M/H] or $Z/X$) is a fixed constant observational constraint, the change in \amlt{} raises both \Xinit{} (by $\approx$0.06) and \Zinit{} (by $\approx$0.004). The increase in fuel and the higher opacity during the main sequence both prolong the main-sequence lifespan, substantially increasing the estimated stellar ages \citep[see also][]{valle++2018-rgb-age}. 
We also note that the sensitivity of \amlt{} is not dependent on whether \Teff{} is being used.

\subsubsection{Helium abundance}

Due to the lack of photospheric helium lines, direct measurements of helium abundance are challenging, leading to \Yinit{} often being treated as a free parameter in stellar modelling. In favorable cases, asteroseismology can measure the helium abundance in the stellar envelope by using the change of the adiabatic index in the HeII ionization zone, which produces oscillatory signatures in p-mode frequencies \citep{basu+2004-solar-helium,broomhall++2014-helium-envelope-rg-kepler,verma++2019-helium,dreau++2020-helium}. 

\rev{Big-bang nucleosynthesis predicts that the primordial helium abundance, $Y_p$, corresponds to a primordial metal abundance of $Z=0$. 
\citet{peimbert++1974} proposed that the chemical evolution of stars in a galaxy follows 
\begin{equation}
    \Yinit{} = Y_p + \frac{\Delta Y}{\Delta Z}\Zinit{},
\end{equation}
where the coefficient $\Delta Y/\Delta Z$ varies with time and location. 
Cosmological models estimate $Y_p$ to range from 0.24 to 0.26, setting a lower boundary on \Yinit{} \citep{planck++2015-primordial-helium}.
In our Galaxy, $\Delta Y/\Delta Z$ is less well-known, with empirical calibrations suggesting values between 1 and 6 \citep{jimenez++2003-He,casagrande++2007-He,verma++2019-helium}.
}

Here, we explore the effects of varying the initial helium abundance \Yinit{}. In our analysis, we modified \Yinit{} from 0.25 to 0.30, aligning with recent seismic estimations \citep{mckkever++2019-ngc6791,ong++2022-helium}. This was implemented by setting priors $\Yinit{}\sim\mathcal{N}(0.25,0.03)$ and $\mathcal{N}(0.30,0.03)$. 
The median differences observed are 0.2\% in mass, 0.1\% in radius, and 9.0\% in age. 
The age discrepancies, although significant, were less pronounced than those caused by changes in \amlt{}, as a result of a smaller range of variation in \Yinit{} (0.05 in this case as opposed to 0.08).
This sensitivity of \Yinit{} is also independent of whether \Teff{} is being used.

\rev{As an example, in Figure~\ref{fig:corner} we show the probability distributions of stellar models for KIC 10904520, constrained by $\chisq{\rm seis}+\chisq{\rm Gaia}+\chisq{\rm spec}$. Both \Yinit{} and \amlt{} exhibit correlations with stellar age. This indicates that the values of stellar ages will be heavily affected by priors on \amlt{} and \Yinit{}. }

\subsubsection{Solar abundance scale}

We also examined the effect of different solar abundance scale $(Z/X)_\odot$ values. The metallicity ratio [M/H], derived from spectroscopy, implicitly sets a specific solar abundance value $(Z/X)_\odot$:
\begin{equation}
    \mh{} = \log_{10}(Z/X) - \log_{10}(Z/X)_\odot.
\end{equation}
Various studies have measured $(Z/X)_\odot$ ranging between 0.0181 and 0.0274 \citep{anders+1989-abundance,grevesse+1998-abundance,asplund++2009-solar-composition-review}. We tested the effects of these variations by altering the model definition of \mh{}, as reflected in \chisq{\rm spec}. The resulting median differences are 0.3\% in mass, 0.1\% in radius, and 10.6\% in age, showing a similar magnitude of impact as the adjustment in \Yinit{} \citep[see also][]{bellinger++2019-errors}.

\subsubsection{Overshoot}

Lastly, we assessed the effect of changing the core overshoot parameter \fov{core}. In stars with masses above $\sim1.2$~\Msolar{}, \fov{core} plays a critical role in determining the convective core boundary and the mixing within. \rev{Empirical calibrations from asteroseismology and eclipsing binaries suggest that \fov{core} increases with stellar mass. Specifically, \fov{core} starts at 0 for stars of $\approx1.2$~\Msolar{} and reaches a stable value of 0.015 for stars above $\approx2.0$~\Msolar{}
\citep{tianzj-2015-subgiant-kic6442183-kic11137075,deheuvels++2016-overshoot,claret++2018-eb-core-fov,mombarg++2021,lindsay++2024-ov,reyes++2024-m67}.}

We compared two typical values for \fov{core} from these studies, using priors $\fov{core}\sim\mathcal{N}(0.005,0.005)$ and $\mathcal{N}(0.015,0.005)$. The changes observed were minimal: negligible in both mass and radius, and 1.0\% in age. We also did not find substantial impacts when changing \fov{shell}, although its value is known to impact the position of the RGB bump \citep{jcd-2015-rgb-bump,khan-2018-rgb-bump-constraints-envelope-overshooting}.

\subsection{Varying the choices of observables}
We can gain further insights into the impact of uncertain input physics by comparing the stellar properties derived from different sets of observables. Figure~\ref{fig:observables} compares the properties derived using various observables against those obtained from the full set: $\chisq{}=\chisq{\rm seis}+\chisq{\rm Gaia}+\chisq{\rm spec}$. 

Our first focus concerns models that incorporate asteroseismic constraints only from individual frequencies: $\nu_{\ell=0}$ and $\nu_{\ell=2}$. These models show discrepancies compared to those using the full set, with systematic offsets of up to 5\% in mass, 1\% in radius, and 15\% in age. We also observed a discernible trend in the age discrepancies, correlating with the age itself. These discrepancies are attributed to the inclusion of \numax{} in the full set, which offers partial constraints on \amlt{} (discussed in Section~\ref{sec:amlt}), leading to a slightly different parameter scale.

We then considered models optimized only by global parameters. We confirmed there are substantial deviations when relying solely on $L$, \Teff{}, and [M/H], as used in traditional isochrone fitting, with systematic offsets of up to 10\% in mass, 2\% in radius, and 30\% in age, compared to those with the full inputs. When incorporating \numax{}, either in addition to or as a replacement for $L$, the discrepancy in radius is similar but the divergence reduces to 5\% in mass and by 15\% in age. This is because \numax{} relates to mass with power of 1 (see Eq.~\ref{eq:sc-numax}). The inclusion of \numax{}, especially for stars beyond $3$~kpc where \Gaia{} distance measurements are less precise, proves to be very beneficial in improving parameter accuracy \citep{huber++2017-seismic-radii-gaia}. 
Furthermore, models incorporating \Dnu{} as an input show excellent consistency with the full seismic inputs. 
Consequently, we recommend employing a combination of \Dnu{} and \numax{} wherever possible, to fully maximize the accuracy of stellar properties \citep{silvaaguirre++2020-tess-rg}.

We compared the masses and ages determined using the full dataset with those determined in earlier studies. 
\citet{wuyq++2018-mass-rgb} optimized Yonsei–Yale (Y2) stellar models \citep{demarque++2004-yy} using \Dnu{}, \numax{}, and spectroscopic parameters from LAMOST spectra. 
The standard deviations of the fraction differences are 7\% for mass and 37\% for age. 
In comparison to \citet{miglio++2021-age-kepler}, who optimized MESA models using \Dnu{}, \numax{}, and spectroscopic parameters from APOGEE spectra,
the standard deviations of the fraction differences are 3\% for mass and 22\% for age. 
Again, these differences highlight the importance of accounting for systematic uncertainties.

\subsection{Summary}

Our investigation reveals that the inherent uncertainties in stellar models can lead to significant deviations in the estimated properties of asteroseismic red giants. Specifically, we observed uncertainties up to $\approx$0.4\% in mass, $\approx$0.2\% in radius and, notably, $\approx$17\% in age.
The uncertainties mainly stem from the poorly-constrained helium abundance (\Yinit{}), which significantly influences other parameters such as the mixing-length parameter (\amlt{}), initial hydrogen fraction (\Xinit{}), and initial metallicity (\Zinit{}). The age uncertainty is particularly critical, often surpassing the statistical uncertainties commonly reported in individual frequency modelling, which are typically around $\approx$10\% \citep{montalban++2021-age,wangyx++2023-rgb}. 
\rev{Subgiants, on the other hand, offer more reliable age scales (\mycitealt{litd++2020-kepler-36-subgiants}; \citealt{tayar++2022-error}; \citealt{xiang++2022}).}

Our results show that, despite integrating constraints from asteroseismic data, a baseline level of systematic uncertainty still persists. This systematic uncertainty sets a realistic lower limit for stellar properties. It is a crucial factor to consider in the application of asteroseismic data for stellar evolution and Galactic archaeology studies.

It is also important to note that, while our study primarily targets RGB stars, the uncertainties for red clump could be even higher due to the accumulation of errors during the earlier evolutionary phases of these stars \citep{cinquegrana++2023-bridge2,noll++2024-rc}.

\begin{figure*}
    \centering
    \includegraphics{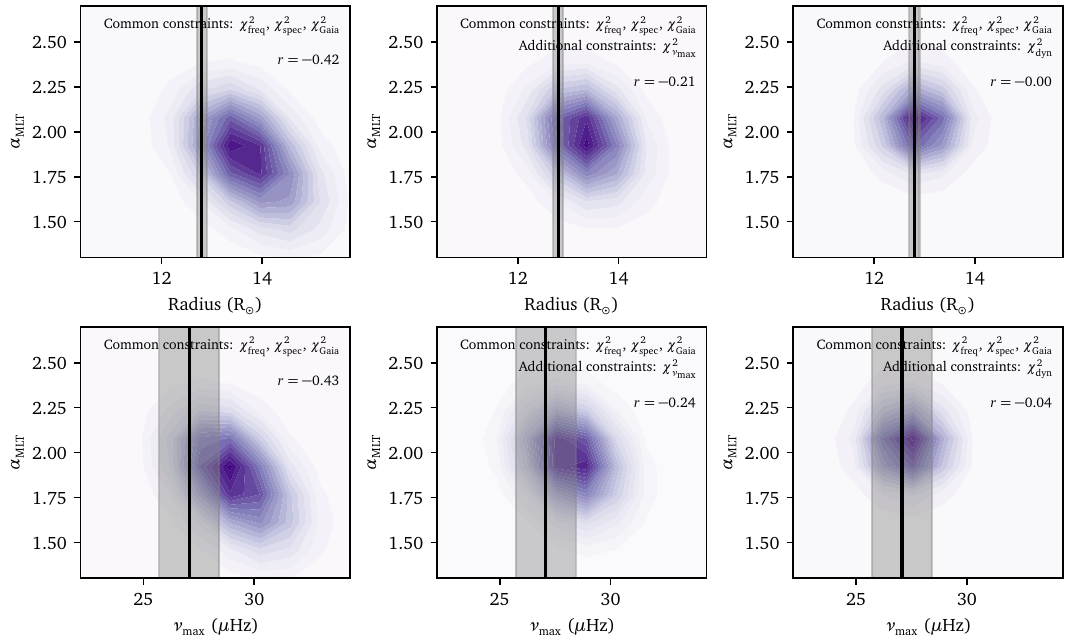}
    \caption{Posterior distributions of stellar models for KIC 9540226 (an eclipsing binary) constrained by $\chisq{}=\chisq{\rm freq}+\chisq{\rm spec}+\chisq{\rm Gaia}$. 
    The top panels show the distributions marginalised in the $(\amlt{}, {\rm radius})$ space and the bottom panels show those in the $(\amlt{}, \numax{})$ space. 
    The panels in the middle column show the results with additional constraint of \chisq{\numax}.
    The panels in the right column show those with additional constraint of \chisq{\rm dyn}.
    The Pearson correlation coefficients calculated from the distributions are displayed in each panel. 
    The vertical lines and the shaded regions show the medians and 1-$\sigma$ values for the measured radius (from binary modelling) and the measured \numax{} (from observed power spectra). }
    \label{fig:corr}
\end{figure*}

\begin{figure}
    \centering
    \includegraphics{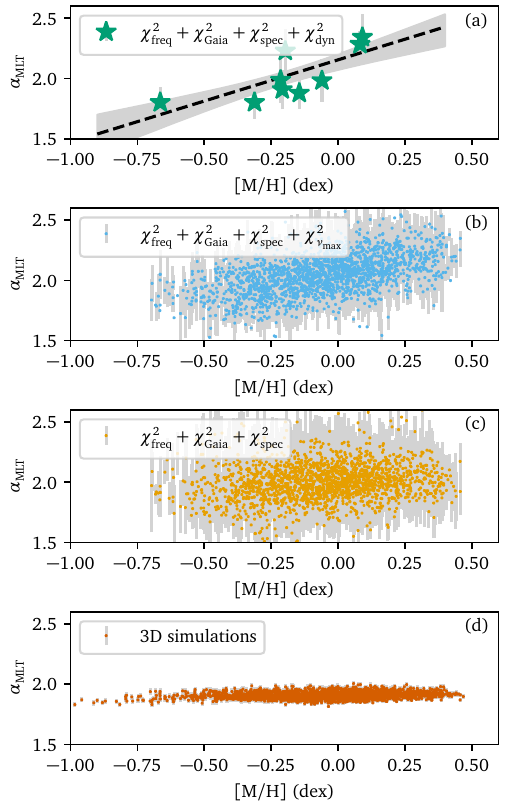}
    \caption{Derived relationships between \amlt{} and metallicity. Panel (a) shows the 1-D modelling results with the additional \chisq{\rm dyn} from the binary sample, panel (b) with \chisq{\numax}, and panel (c) with neither of them. Panel (d) shows the interpolated \amlt{} from 3-D simulations at measured \Teff{}, [M/H] and \logg{}. }
    \label{fig:amlt}
\end{figure}

\begin{figure*}
    \centering
    \includegraphics{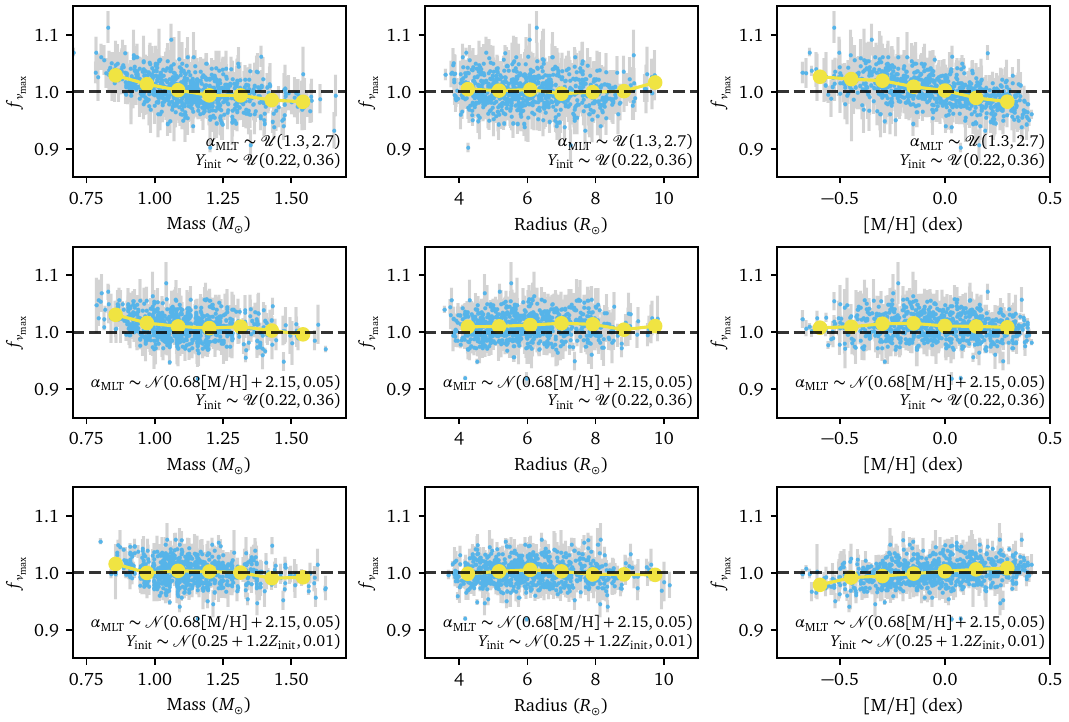}
    \caption{Correction factor of the \numax{} scaling relation, \fnumax{}, as a function of mass, radius, and metallicity (from left to right panels). The top rows show the results obtained without imposing informative priors on \amlt{} and \Yinit{}. The middle rows show those with the calibrated \amlt{}--[M/H] relation from the eclipsing binary sample. The bottom rows show those with an assumed helium enrichment law.}
    \label{fig:fnumax}
\end{figure*}

\section{Constraining the mixing-length parameter}\label{sec:amlt}

\subsection{Interplay between \texorpdfstring{\amlt{}}{the mixing-length parameter}, radius, and \texorpdfstring{\numax{}}{numax}}

Our analysis of stellar models, when we omit \numax{} as a constraint and rely on $\chisq{} = \chisq{\rm freq} + \chisq{\rm Gaia} + \chisq{\rm spec}$, reveals a notable challenge in precisely determining stellar the radius. This difficulty stems from the lack of constraint on the mixing-length parameter (\amlt{}). We explain this issue in greater detail below.

In the upper-left panel of Figure~\ref{fig:corr}, we present the posterior distribution of KIC 9540226, marginalised over the \amlt{}--radius parameter space. It reveals a strong negative and nearly linear correlation between the two parameters under the observational constraints, with a Pearson correlation coefficient of $-0.43$. The correlation suggests that accurate determination of either \amlt{} or radius is challenging without incorporating additional data.

In the upper-middle panel of Figure~\ref{fig:corr}, we introduce an extra constraint, \numax{}. Since our 1-D models do not predict amplitudes, the \numax{} values were calculated from the scaling relation:
\begin{equation}\label{eq:sc-numax}
\frac{\nu_{\rm max}}{\nu_{\rm max,\odot}} \approx \left(\frac{M}{M_{\odot}}\right) \left(\frac{R}{R_{\odot}}\right)^{-2} \left(\frac{T_{\rm eff}}{T_{\rm eff,\odot}}\right)^{-1/2},
\end{equation}
where $\nu_{{\text{max}}, \odot}=3090$ \muHz{}, and $T_{{\text{eff}}, \odot}=5772$ K.
This inclusion visibly weakens the correlation (Pearson correlation coefficient reduced to $-0.21$) because \numax{} is proportional to $MR^{-2}T^{-1/2}_{\rm eff}$, indirectly imposing a constraint on the radius.

In the upper-right panel of Figure~\ref{fig:corr}, we applied additional constraints, \chisq{\rm dyn}, related to the mass and radius of KIC 9540266. This effectively neutralizes the correlation between \amlt{} and radius, allowing for a precise determination of \amlt{}. \citet{joyce+2018-alpha-cen} applied a similar approach to the $\alpha$ Cen A \& B binary system, demonstrating a relation between \amlt{} and stellar mass that was preserved across all variations in input physics. We also tested with the radius constraint alone, which yielded similar outcomes. This indicates the importance of direct radius measurements, such as those from interferometry and binary orbit modelling, for calibrating \amlt{} in stars beyond the Sun.

In the lower panels of Figure~\ref{fig:corr}, we present the posterior distributions now marginalised over the \amlt{}--\numax{} space. 
We found that precise \numax{} determination from models hinges on having \amlt{} calibrated with direct stellar radius data. 
It highlights an important point: predicting a reliable \numax{} value based solely on individual frequencies is nearly impractical without a predetermined \amlt{}. 
This is because the individual frequencies are mostly sensitive to deeper regions within the star, and their sensitivity to super adiabatic near-surface regions is considerably reduced by the empirical surface correction procedure.

\subsection{Relation between \texorpdfstring{\amlt{}}{the mixing-length parameter} and metallicity}\label{subsec:amlt-mh}

We now turn to examining the relationship between \amlt{} and [M/H]. It has been shown in previous studies (such as \citealt{metcalfe++2014-amp,creevey++2017-amp,tayar-2017-amlt,joyce+2018-amlt-metal-poor,viani++2018-amlt-mh}; \mycitealt{litd++2018-amlt}; \citealt{valle++2019-amlt}) that 1-D stellar models, when constrained by asteroseismic data, often require adjustments in the \amlt{} values depending the metallicity. This relationship is typically found as a monotonically increasing trend between the two parameters.

We now revisit the \amlt{}--[M/H] relation, which we intend to constrain with the dynamical mass and radius \chisq{\rm dyn}. The results are shown in Figure~\ref{fig:amlt}. 
With the \chisq{\rm dyn} constraints (panel a), we still observe a positive correlation between \amlt{} and [M/H], akin to the pattern found in previous 1-D studies and in models constrained by \chisq{\numax{}} (panel b). 
However, there is a slight difference in the correlation slopes when comparing the two methods. This variation can be attributed to the different effectiveness of \chisq{\numax{}} and \chisq{\rm dyn} in mitigating the correlation between \amlt{} and radius, as previously discussed.
When neither \chisq{\rm dyn} nor \chisq{\numax{}} are applied (panel c), the correlation between \amlt{} and [M/H] almost vanishes. 
Therefore, we highlight the importance of having effective calibrations on \amlt{} for model-based parameter inference. Relying solely on models without such priors could lead to a systematic bias in the derived stellar properties, as we discussed in Section~\ref{sec:property}.

We also checked the \amlt{} parameter informed by the Stagger models, a grid of 3-D hydrodynamic simulations of stellar surfaces \citep{magic++2013-stagger-1}. 
\citet{magic++2015-stagger-3-amlt} calibrated \amlt{} values by matching the entropy values in the nearly adiabatic convection zone of 3-D models with those from corresponding 1-D model atmospheres. 
We determined the \amlt{} of our sample by interpolating their calibrated \amlt{} values at the observed \Teff{}, [M/H] and \logg{} for our sample. 
The results are shown in Figure~\ref{fig:amlt}d, where we do not observe a steep metallicity gradient as seen from 1-D stellar modelling. 
This discrepancy has been acknowledged by \citet{tayar-2017-amlt} and \citet{viani++2018-amlt-mh} and the reason is not entirely clear. 
One possibility is that the helium mass fraction $Y$ increases with decreasing metallicity in the Stagger-grid models, which contradicts the Galactic helium enrichment law. Lowering $Y$ in metal-poor 3-D models gives a smaller overall density \citep{Karlsmose-thesis}, which has a similar effect as lowering the surface gravity hence likely to result in smaller calibrated \amlt{} \citep{trampedach++2014-amlt,magic++2015-stagger-3-amlt}. Had the 3-D simulations constructed following the Helium enrichment law, the expectation is that the positive correlation between [M/H] and calibrated \amlt{} might be more pronounced.

Using the mass and radius measurements from the eclipsing binary sample, along with the spectroscopic and asteroseismic constraints, we fitted a linear relation between \amlt{} and [M/H]: 
\begin{equation}\label{eq:amlt-mh}
    \amlt = (0.66 \pm 0.20) \times {\rm [M/H]} + (2.14 \pm 0.06).
\end{equation}
We verified these results by replacing \Teff{} values from those from APOGEE, and found a similar relation:
\begin{equation}\label{eq:amlt-mh-apogee}
    \amlt = (0.60 \pm 0.14) \times {\rm [M/H]} + (2.14 \pm 0.06).
\end{equation}
The slope of \amlt{} with respect to \mh{} agrees well with the findings of \citet{viani++2018-amlt-mh}, who modelled stars with \Dnu{}, \numax{}, \Teff{}, and \mh{} and reported a slope of $\approx$0.74. However, our value appears to be higher than the $\approx$0.16 value found by \citet{tayar-2017-amlt}, which might be due to their focus on matching \Teff{} instead.

\rev{Although we rely on the binary sample to calibrate Eq.~\ref{eq:amlt-mh}, we caution that binary stars may be subject to radius inflation. It has been discovered in K and M main-sequence stars that the observed stellar radii are systematically larger than predicted by theoretical models \citep{torres++2002-eb}. Magnetic fields \citep{feiden++2014} and starspots \citep{jackson++2014,somers++2015} may be responsible for this discrepancy. Since many red giants in binary systems are magnetically active \citep{gaulme++20-active-rg,gehan++2024-rg-activity}, it is reasonable to conjecture that the radii of these red giant binaries may also be affected by similar issues. }

We investigated whether \mh{} is the primary driver for the variability of \amlt{} in red giants. We calculated the correlation coefficients of \amlt{} with respect to \mh{}, mass, and \Teff{} for the binary sample. They are 0.75, 0.74, and 0.70, respectively, suggesting that a relation with mass or \Teff{} could be equally plausible. However, we have to keep in mind that the binary sample only spans a rather small parameter range. In contrast, for the large asteroseismic sample constrained by \numax{}, we found those coefficients to be 0.46, 0.10, and 0.02, respectively. Assuming this \numax{}-constrained scenario is accurate, then \mh{} is indeed the dominant factor influencing variations in \amlt{} for red giants.

Based on these findings, we adopt Eq.~\ref{eq:amlt-mh} as a preliminary model for further discussions in the subsequent section. We encourage future studies to further refine and validate this relationship as more data from similar systems become available. In particular, more calibrators at lower luminosities and higher metallicities will be especially helpful (see Figure~\ref{fig:hrd}).

\section{Implications on testing the \texorpdfstring{\numax{}}{numax} scaling relation}\label{sec:numax}

\subsection{Background}

The \numax{} scaling relation is an extremely useful tool for measuring stellar parameters, especially in the current era of extensive ensemble analysis of red giants. 
Despite its widespread use, the limitations and full scope of this relation are not completely understood \citep{belkacem++2011-physics-under-numax-nuc,hekker-2020-scaling-review}. 
In this section, we discuss the test of this scaling relation using 1-D stellar modelling. 

The frequency of maximum power, \numax{}, is governed by excitation and damping mechanisms, and is therefore inherently a surface property. In this sense, it is different from the individual modes frequencies, which are the resonant frequencies of the entire star. Given that solar-like oscillations are driven and damped by near-surface convection \citep{goldreich++1994-excitation-solar-pmodes,samadi++2012-amplitudes-rg-corot,zhouyx++2020-rg}, it is reasonable to suppose that \numax{}, like amplitude, is determined by the surface properties of stars, namely effective temperature, surface gravity and possibly metallicity.

The \numax{} scaling relation, originally proposed for main-sequence stars by \citet{brown++1991-dection-procyon-scaling-relation} and \citet{kjeldsen+1995-scaling-relations}, posits that \numax{} is a fixed fraction of the acoustic cutoff frequency (\nuac{}) in the stellar atmosphere, which sets the upper boundary limit for sound waves to remain trapped within the star. 
Under assumptions of an isothermal atmosphere and ideal gas conditions\footnote{These are simplifications that, admittedly, aren't strictly accurate.}, \nuac{} can be shown to scale as $\nuac\propto g/\sqrt{T_{\rm eff}}$, where $g$ is the surface gravity and \Teff{} is the effective temperature. 
To quantify deviation from the scaling relation, we can define a factor \fnumax{} as follows \citep{sharma++2016-population-rg-kepler}:
\begin{equation}\label{eq:fnumax}
\frac{\nu_{\rm max}}{\nu_{\rm max,\odot}} = f_{\numax{}} \left(\frac{M}{M_{\odot}}\right) \left(\frac{R}{R_{\odot}}\right)^{-2} \left(\frac{T_{\rm eff}}{T_{\rm eff,\odot}}\right)^{-1/2}.
\end{equation}

The \numax{} scaling relation has undergone extensive validation against various methodologies, such as astrometry-based luminosities \citep{silvaaguirre++2012-seismic-parallax,huber++2017-seismic-radii-gaia,sahlholdt+2018-gaiadr2-sc-radius-dwarfs,hall++2019-rc-gaiadr2-seismo,khan++2019-gaiadr2-zero-point,zinn++2019-radius-sc}, masses and radii from eclipsing binaries \citep{gaulme++2016-eb-sc,brogaard++2018-accuracy-scaling-relation,kallinger++2018-sc,benbakoura++2021-binary}, and the sharpness of population-level features, such as the zero-age main-sequence edge in red giants \mycitep{liyg++2021-sc-intrinsic-scatter}.

Two interesting deviations have emerged in the extremes of parameter space. First, there is a noticeable deviation for red giants with radii larger than $\approx$50 \Rsolar{}, of which the reason is not entirely clear \citep{yuj++2020-lpv1,zinn++2023-adiabatic}. 
Second, there is a potentially unaccounted-for dependency of \numax{} on metallicity, supported by possible correlations seen in model-based stellar parameters (\mycitealt{litd++2022-kepler-rgb}; \citealt{wangyx++2023-rgb}), and the tendency of \numax{}-based stellar mass to be overestimated in metal-poor stars \citep{epstein++2014-sc-metal-poor,schonhut-stasik++2023-k2}. 
An additional metallicity term could resolve these discrepancies.

Indeed, by using a more accurate derivation for sound speed, \citet{viani++2017-numax-sc-metal} proposed that \nuac{} can also depend on the mean molecular weight and the adiabatic index, thereby introducing a metallicity component \citep[see also][]{jimenez++2015-nuac-six-kepler,yildiz++2016-scaling-relation-tuning}. 
However, recent 3-D hydrodynamic atmosphere simulations by \citet{zhouyx++2023-numax} did not reveal significant variations of \numax{} with metallicity, which could be due to a counteracting effect from the Mach number dependency, as suggested by \citet{belkacem++2011-physics-under-numax-nuc}.

\subsection{Results and discussions}~\label{subsec:numax-results}

We tested the \numax{} scaling relation using the individual frequency modelling approach, similar to prior studies including \citet{metcalfe++2014-amp}, \citet{coelho++2015-numax}, \mycitet{litd++2022-kepler-rgb} and \citet{wangyx++2023-rgb}. 
Our approach involves computing a scaling \numax{} according to Eq.~\ref{eq:sc-numax}, using observational properties other than \numax{} as constraints: $\chisq{}=\chisq{\rm freq} + \chisq{\rm Gaia} + \chisq{\rm spec}$. 
The derived scaling \numax{} is then compared against the actual measured \numax{} from power spectra. 
The ratio of the measured to the derived value is \fnumax{} (see Eq.~\ref{eq:fnumax}). 
Our method differs from previous studies in two aspects: (1) we incorporate a more comprehensive list of observational constraints including $\ell=0-2$ mode frequencies, to get more precise constraints on the internal structure; (2) we study how model uncertainties affect the derived \numax{}.

Figure~\ref{fig:fnumax} presents the correction factor \fnumax{} as a function of mass, radius, and age. 
We also employ median values aggregated into bins (displayed as circles), to highlight any systematic trends in \fnumax{}.
Our initial analysis, presented in the top row of the figure, was conducted without applying informative priors on \amlt{} and \Yinit{}. 
These results indicate a negative correlation of \fnumax{} with both mass and metallicity.
Such findings could lead to the conclusions that adjustments to the \numax{} scaling relation are necessary. 
The correlation with [M/H] is especially tempting, considering the previously discussed influence of [M/H].

However, it is essential to approach these trends cautiously. 
Our previous discussions suggest that \amlt{} is not adequately constrained even with asteroseismic frequencies. Thus, employing non-informative priors on \amlt{} can result in systematically biased stellar parameters. 
To address this, we used a normal prior on \amlt{} around the value given by Eq.~\ref{eq:amlt-mh}:
\begin{equation}
    \amlt{} \sim {\mathcal{N} (0.68{\rm [M/H]} + 2.15, 0.05)}.
\end{equation}
The resulting \fnumax{} is shown in the middle row of Figure~\ref{fig:fnumax}. 
The trend with [M/H] vanishes when we incorporate the calibrated \amlt{}-[M/H] relation. 
The trend with mass also greatly reduces.

Interestingly in this case, we noticed that the \fnumax{} factor consistently exceeds 1, hinting at a possible need to adjust the reference values in Eq.~\ref{eq:sc-numax} for red giants from the current solar-based values. 
Yet, this adjustment may not be necessary if we apply specific priors to \Yinit{}.
We implemented a normal prior for \Yinit{} based on the Galactic enrichment law:
\begin{equation}
    \Yinit{} \sim {\mathcal{N} (Y_p + \frac{\Delta Y}{\Delta Z}\Zinit{}, 0.01)},
\end{equation}
with $Y_p = 0.25$ and ${\Delta Y}/{\Delta Z} = 1.2$, although the value for ${\Delta Y}/{\Delta Z}$ is subject to debate. 
The outcomes of this implementation are shown in the bottom row of Figure~\ref{fig:fnumax}. 
We observed that the \fnumax{} correction factor closely aligns with 1, suggesting no significant deviation from the standard scaling relation. 
We also note that in this case \fnumax{} presents a positive correlation with [M/H], which can be easily destructed again with a slightly different prescription for \amlt{}.

From these findings, we conclude that testing the \numax{} scaling relation through individual frequency modelling does not always yield conclusive results and requires a thorough consideration of model uncertainties, especially \amlt{} and \Yinit{}. 
Additionally, we may also expect that other model assumptions, such as the atmosphere boundary condition \citep{Choi++2018-rgb-boundary-condition-100K-teff-uncertain,salaris++2018-teff-rg-apokasc-mixing-length-calibration} and the treatment of surface correction \mycitep{liyg++2022-surface}, modify the surface layers and influence the derived scaling \numax{}.

Our findings also indicate that \numax{} imparts unique information distinct from individual frequencies. Therefore, \numax{} should be regarded as a valuable constraint in asteroseismic modelling. 

\subsection{Recommendations for using the \texorpdfstring{\numax{}}{numax} scaling relation}\label{subsec:recommendation}

We recommend practices for using \numax{} to derive stellar properties. Most \numax{} extraction pipelines only consider the statistical uncertainties for measuring \numax{}, which sometimes could be smaller than the total error budget. Below we list some of the external sources of uncertainties. 

First, the values of \numax{} is influenced by the method in which stellar oscillation was detected. \numax{} is affected by the wavelength range of the photometric filters used, as well as whether observations are conducted through radial velocity or photometry. Such factors can result in $\approx$5\% discrepancies in \numax{} in the Sun \citep{howe++2020-numax-solar}. Moreover, differences in the treatment of power spectra among data reduction pipelines can lead to differences in \numax{} for about 2\% \citep{pinsonneault++2018-apokasc}. Therefore, when the studied samples are not characterised with the same instrument or the same data reduction pipeline, it is crucial to account for these additional sources of discrepancy.

Second, \numax{} changes over time due to astrophysical origins. For example, \numax{} is subject to variations with magnetic cycles. The Sun varies \numax{} of $\approx$0.8\% in a solar cycle \citep{howe++2020-numax-solar}. In addition, the stochastic nature of oscillations can cause \numax{} to fluctuate over fixed lengths of observing windows. In the case of \Kepler{} red giants, typical scatter in \numax{} ranges from 2\% to 5\% in three-month observing windows \citep{sreenivas++2024-numax}. This issue is even more troublesome for some TESS stars with only one sector of data \citep[e.g.][]{jiangc++2023-hd76920}. Hence, these types of uncertainties need to be quantified, presumably via simulations, especially if \numax{} is measured with short time series.

After accounting for the various sources of uncertainty associated with \numax{}, we advocate for its integration into stellar modelling. 
As detailed in Section~\ref{subsec:numax-results}, \numax{} provides constraints distinct from those obtained from individual frequencies.
Furthermore, as illustrated in Section~\ref{subsec:amlt-mh}, incorporating \numax{} assists in reducing the uncertainty in the mixing-length parameter, \amlt{}. This reduction is especially helpful for reducing uncertainties in stellar age, which arises from mutual effects of \amlt{} and \Yinit{}  (Section~\ref{sec:property}).

\section{Conclusions}

In our study, we present a detailed characterisation of red giant model uncertainties, focusing on the impact of the mixing-length parameter \amlt{}, the initial helium fraction \Yinit{}, the solar abundance scale $(Z/X)_\odot$, and the core overshoot parameter \fov{core}. Our objective is to understand how these factors influence the determination of stellar mass, radius, and age in asteroseismic modelling. The key conclusions are summarised as follows:
\begin{enumerate}
    \item We identified that uncertainties in \amlt{} and \Yinit{} significantly affect the accuracy of stellar properties, despite incorporating constraints from spectroscopy (\Teff{}, [M/H]), \Gaia{} astrometry ($L$), and asteroseismology ($\nu_{\ell=0}$, $\nu_{\ell=2}$, \numax{}, \DPi{1}). These uncertainties set an error floor on mass of $\approx$0.4\%, radius of $\approx$0.2\%, and age of $\approx$17\% (Figure~\ref{fig:physics}). The error floors due to model uncertainties in age exceed typical statistical uncertainties $\approx$10\%, showing the importance of their evaluation in asteroseismic applications. 
    \item We examined the effect of different combinations of observational constraints (Figure~\ref{fig:observables}). Incorporating asteroseismic constraints only from individual frequencies shows discrepancies compared to the case of using the full set of inputs that include \numax{}, suggesting \numax{} sets a slightly different parameter scale, due to its constraining ability on \amlt{}. 
    In addition, incorporating asteroseismic constraints from \Dnu{} and \numax{} shows excellent consistency with the full set of inputs. Therefore, for ensemble analysis of red giants, a combination of \Dnu{} and \numax{} should be included wherever possible. 
    \item We showed that an uncertain state of \amlt{} translates to uncertain predictions on stellar radius and \numax{} (Figure~\ref{fig:corr}). Hence, direct and accurate measurements of stellar radius will enable the calibration of \amlt{}, such as those from eclipsing binaries or interferometry. More calibrators are needed in order to fully address this uncertainty.
    \item Using a small \Kepler{} eclipsing binary sample, for which dynamical radii have been determined, we calibrated the relation between \amlt{} and [M/H] (Eq.~\ref{eq:amlt-mh} and Figure~\ref{fig:amlt}). We observed a positive correlation, consistent with previous 1-D stellar asteroseismic modelling that does not use the extra constraints from binaries. The result remains in tension with 3-D simulations, possibly due to differing treatments of helium abundances.
    \item We showed that predicting a reliable \numax{} value based solely on individual frequencies is nearly impractical (Figure~\ref{fig:corr}). The model-derived scaling \numax{}, shows strong dependence on the uncertainties in both \amlt{} and \Yinit{}, which prohibits accuracy tests on the \numax{} scaling relation from mode frequency modelling (Figure~\ref{fig:fnumax}). 
    \item We concluded that \numax{} provides distinct information and should be considered as an important observable in asteroseismic modelling, and provided our guidelines for its usage in Section~\ref{subsec:recommendation}.
\end{enumerate}

\section*{acknowledgements}
Y.L. expresses gratitude to Bill Chaplin and Margarida Cunha for their thorough and constructive comments during his PhD exam process.
Y.L. acknowledges the support by the Beatrice Watson Parrent Fellowship and the National Aeronautics and Space Administration (80NSSC19K0597).
We acknowledge support from the Australian Research Council for T.R.B (DP210103119 and FL220100117), D.S. (DP190100666) and S.J.M (FT210100485).
D.H. acknowledges support from the Alfred P. Sloan Foundation and the Australian Research Council (FT200100871).
M.J. is supported by the Horizon 2020 research and innovation programme's funding of MATISSE: \textit{Measuring Ages Through Isochrones, Seismology, and Stellar Evolution}.

Funding for the Kepler mission is provided by the NASA Science Mission Directorate. This paper includes data collected by the Kepler mission and obtained from the MAST data archive at the Space Telescope Science Institute (STScI). STScI is operated by the Association of Universities for Research in Astronomy, Inc., under NASA contract NAS 5–26555.


This work presents results from the European Space Agency (ESA) space mission Gaia. Gaia data are being processed by the Gaia Data Processing and Analysis Consortium (DPAC). Funding for the DPAC is provided by national institutions, in particular the institutions participating in the Gaia MultiLateral Agreement (MLA). 

The APOGEE data is from the Sloan Digital Sky Survey IV, whose funding has been provided by the Alfred P. Sloan Foundation, the U.S. Department of Energy Office of Science, and the Participating Institutions. 

We acknowledge the Sydney Informatics, the University of Sydney’s high performance computing (HPC) cluster Artemis, and the University of Hawaii's Information Technology Services – Cyberinfrastructure (funded in part by the National Science Foundation CC* awards \#2201428 and \#2232862), for providing the HPC resources that have contributed to the research results reported within this paper.


{\emph{Software}}: {Numpy} \citep{numpy}, {Scipy} \citep{scipy}, {Matplotlib} \citep{matplotlib}, {Astropy} \citep{astropy1,astropy2,astropy3}, {Pandas} \citep{pandas}, {MESA} \citep{paxton++2011-mesa,paxton++2013-mesa,paxton++2015-mesa,paxton++2018-mesa,paxton++2019-mesa,jermyn++2023-mesa}, {MESASDK} \citep{mesasdk}, {GYRE} \citep{townsend+2013-gyre}, {Lightkurve} \citep{lightkurve}, {ISOCLASSIFY} \citep{huber++2017-seismic-radii-gaia,berger++2020-gaia-kepler-1-stars}, {colte} \citep{casagrande++2021-irfm-gaia}, {GPT-4} \citep{gpt4}.
\rev{The analysis scripts, MESA inlists and MESA models used in this work are available at GitHub (link) \footnote{insert Github link upon publication.} and Zenodo (cite DOI)}.

\bibliography{references/myastrobib}{}

\begin{thebibliography}{154}
\expandafter\ifx\csname natexlab\endcsname\relax\def\natexlab#1{#1}\fi

\bibitem[{{Abdurro'uf} {et~al.}(2022){Abdurro'uf}, {Accetta}, {Aerts}, {Silva
  Aguirre}, {Ahumada}, {Ajgaonkar}, {Filiz Ak}, {Alam}, {Allende Prieto},
  {Almeida}, {Anders}, {Anderson}, {Andrews}, {Anguiano}, {Aquino-Ort{\'\i}z},
  {Arag{\'o}n-Salamanca}, {Argudo-Fern{\'a}ndez}, {Ata}, {Aubert},
  {Avila-Reese}, {Badenes}, {Barb{\'a}}, {Barger}, {Barrera-Ballesteros},
  {Beaton}, {Beers}, {Belfiore}, {Bender}, {Bernardi}, {Bershady}, {Beutler},
  {Bidin}, {Bird}, {Bizyaev}, {Blanc}, {Blanton}, {Boardman}, {Bolton},
  {Boquien}, {Borissova}, {Bovy}, {Brandt}, {Brown}, {Brownstein}, {Brusa},
  {Buchner}, {Bundy}, {Burchett}, {Bureau}, {Burgasser}, {Cabang}, {Campbell},
  {Cappellari}, {Carlberg}, {Wanderley}, {Carrera}, {Cash}, {Chen}, {Chen},
  {Cherinka}, {Chiappini}, {Choi}, {Chojnowski}, {Chung}, {Clerc}, {Cohen},
  {Comerford}, {Comparat}, {da Costa}, {Covey}, {Crane}, {Cruz-Gonzalez},
  {Culhane}, {Cunha}, {Dai}, {Damke}, {Darling}, {Davidson}, {Davies},
  {Dawson}, {De Lee}, {Diamond-Stanic}, {Cano-D{\'\i}az}, {S{\'a}nchez},
  {Donor}, {Duckworth}, {Dwelly}, {Eisenstein}, {Elsworth}, {Emsellem},
  {Eracleous}, {Escoffier}, {Fan}, {Farr}, {Feng}, {Fern{\'a}ndez-Trincado},
  {Feuillet}, {Filipp}, {Fillingham}, {Frinchaboy}, {Fromenteau}, {Galbany},
  {Garc{\'\i}a}, {Garc{\'\i}a-Hern{\'a}ndez}, {Ge}, {Geisler}, {Gelfand},
  {G{\'e}ron}, {Gibson}, {Goddy}, {Godoy-Rivera}, {Grabowski}, {Green},
  {Greener}, {Grier}, {Griffith}, {Guo}, {Guy}, {Hadjara}, {Harding},
  {Hasselquist}, {Hayes}, {Hearty}, {Hern{\'a}ndez}, {Hill}, {Hogg},
  {Holtzman}, {Horta}, {Hsieh}, {Hsu}, {Hsu}, {Huber}, {Huertas-Company},
  {Hutchinson}, {Hwang}, {Ibarra-Medel}, {Chitham}, {Ilha}, {Imig}, {Jaekle},
  {Jayasinghe}, {Ji}, {Johnson}, {Jones}, {J{\"o}nsson}, {Katkov}, {Khalatyan},
  {Kinemuchi}, {Kisku}, {Knapen}, {Kneib}, {Kollmeier}, {Kong}, {Kounkel},
  {Kreckel}, {Krishnarao}, {Lacerna}, {Lane}, {Langgin}, {Lavender}, {Law},
  {Lazarz}, {Leung}, {Leung}, {Lewis}, {Li}, {Li}, {Lian}, {Liang}, {Lin},
  {Lin}, {Lin}, {Lintott}, {Long}, {Longa-Pe{\~n}a}, {L{\'o}pez-Cob{\'a}},
  {Lu}, {Lundgren}, {Luo}, {Mackereth}, {de la Macorra}, {Mahadevan},
  {Majewski}, {Manchado}, {Mandeville}, {Maraston}, {Margalef-Bentabol},
  {Masseron}, {Masters}, {Mathur}, {McDermid}, {Mckay}, {Merloni},
  {Merrifield}, {Meszaros}, {Miglio}, {Di Mille}, {Minniti}, {Minsley},
  {Monachesi}, {Moon}, {Mosser}, {Mulchaey}, {Muna}, {Mu{\~n}oz}, {Myers},
  {Myers}, {Nadathur}, {Nair}, {Nandra}, {Neumann}, {Newman}, {Nidever},
  {Nikakhtar}, {Nitschelm}, {O'Connell}, {Garma-Oehmichen}, {Luan Souza de
  Oliveira}, {Olney}, {Oravetz}, {Ortigoza-Urdaneta}, {Osorio}, {Otter},
  {Pace}, {Padilla}, {Pan}, {Pan}, {Parikh}, {Parker}, {Peirani}, {Pe{\~n}a
  Ram{\'\i}rez}, {Penny}, {Percival}, {Perez-Fournon}, {Pinsonneault},
  {Poidevin}, {Poovelil}, {Price-Whelan}, {B{\'a}rbara de Andrade Queiroz},
  {Raddick}, {Ray}, {Rembold}, {Riddle}, {Riffel}, {Riffel}, {Rix}, {Robin},
  {Rodr{\'\i}guez-Puebla}, {Roman-Lopes}, {Rom{\'a}n-Z{\'u}{\~n}iga}, {Rose},
  {Ross}, {Rossi}, {Rubin}, {Salvato}, {S{\'a}nchez}, {S{\'a}nchez-Gallego},
  {Sanderson}, {Santana Rojas}, {Sarceno}, {Sarmiento}, {Sayres}, {Sazonova},
  {Schaefer}, {Schiavon}, {Schlegel}, {Schneider}, {Schultheis}, {Schwope},
  {Serenelli}, {Serna}, {Shao}, {Shapiro}, {Sharma}, {Shen}, {Shetrone}, {Shu},
  {Simon}, {Skrutskie}, {Smethurst}, {Smith}, {Sobeck}, {Spoo}, {Sprague},
  {Stark}, {Stassun}, {Steinmetz}, {Stello}, {Stone-Martinez},
  {Storchi-Bergmann}, {Stringfellow}, {Stutz}, {Su}, {Taghizadeh-Popp},
  {Talbot}, {Tayar}, {Telles}, {Teske}, {Thakar}, {Theissen}, {Tkachenko},
  {Thomas}, {Tojeiro}, {Hernandez Toledo}, {Troup}, {Trump}, {Trussler},
  {Turner}, {Tuttle}, {Unda-Sanzana}, {V{\'a}zquez-Mata}, {Valentini},
  {Valenzuela}, {Vargas-Gonz{\'a}lez}, {Vargas-Maga{\~n}a}, {Alfaro},
  {Villanova}, {Vincenzo}, {Wake}, {Warfield}, {Washington}, {Weaver},
  {Weijmans}, {Weinberg}, {Weiss}, {Westfall}, {Wild}, {Wilde}, {Wilson},
  {Wilson}, {Wilson}, {Wolf}, {Wood-Vasey}, {Yan}, {Zamora}, {Zasowski},
  {Zhang}, {Zhao}, {Zheng}, {Zheng}, \& {Zhu}}]{abdurrouf++2021-apogee-dr17}
{Abdurro'uf}, {Accetta}, K., {Aerts}, C., {et~al.} 2022, \apjs, 259, 35

\bibitem[{{Anders} \& {Grevesse}(1989)}]{anders+1989-abundance}
{Anders}, E., \& {Grevesse}, N. 1989, \gca, 53, 197

\bibitem[{{Anders} {et~al.}(2023){Anders}, {Gispert}, {Ratcliffe}, {Chiappini},
  {Minchev}, {Nepal}, {Queiroz}, {Amarante}, {Antoja}, {Casali}, {Casamiquela},
  {Khalatyan}, {Miglio}, {Perottoni}, \&
  {Schultheis}}]{anders++2023-apogee-age}
{Anders}, F., {Gispert}, P., {Ratcliffe}, B., {et~al.} 2023, \aap, 678, A158

\bibitem[{{Appourchaux} {et~al.}(2015){Appourchaux}, {Antia}, {Ball},
  {Creevey}, {Lebreton}, {Verma}, {Vorontsov}, {Campante}, {Davies}, {Gaulme},
  {R{\'e}gulo}, {Horch}, {Howell}, {Everett}, {Ciardi}, {Fossati}, {Miglio},
  {Montalb{\'a}n}, {Chaplin}, {Garc{\'\i}a}, \&
  {Gizon}}]{appouchaux++2015-hip93511-binary}
{Appourchaux}, T., {Antia}, H.~M., {Ball}, W., {et~al.} 2015, \aap, 582, A25

\bibitem[{{Asplund} {et~al.}(2009){Asplund}, {Grevesse}, {Sauval}, \&
  {Scott}}]{asplund++2009-solar-composition-review}
{Asplund}, M., {Grevesse}, N., {Sauval}, A.~J., \& {Scott}, P. 2009, \araa, 47,
  481

\bibitem[{{Astropy Collaboration} {et~al.}(2013){Astropy Collaboration},
  {Robitaille}, {Tollerud}, {Greenfield}, {Droettboom}, {Bray}, {Aldcroft},
  {Davis}, {Ginsburg}, {Price-Whelan}, {Kerzendorf}, {Conley}, {Crighton},
  {Barbary}, {Muna}, {Ferguson}, {Grollier}, {Parikh}, {Nair}, {Unther},
  {Deil}, {Woillez}, {Conseil}, {Kramer}, {Turner}, {Singer}, {Fox}, {Weaver},
  {Zabalza}, {Edwards}, {Azalee Bostroem}, {Burke}, {Casey}, {Crawford},
  {Dencheva}, {Ely}, {Jenness}, {Labrie}, {Lim}, {Pierfederici}, {Pontzen},
  {Ptak}, {Refsdal}, {Servillat}, \& {Streicher}}]{astropy1}
{Astropy Collaboration}, {Robitaille}, T.~P., {Tollerud}, E.~J., {et~al.} 2013,
  \aap, 558, A33

\bibitem[{{Astropy Collaboration} {et~al.}(2018){Astropy Collaboration},
  {Price-Whelan}, {Sip{\H{o}}cz}, {G{\"u}nther}, {Lim}, {Crawford}, {Conseil},
  {Shupe}, {Craig}, {Dencheva}, {Ginsburg}, {VanderPlas}, {Bradley},
  {P{\'e}rez-Su{\'a}rez}, {de Val-Borro}, {Aldcroft}, {Cruz}, {Robitaille},
  {Tollerud}, {Ardelean}, {Babej}, {Bach}, {Bachetti}, {Bakanov}, {Bamford},
  {Barentsen}, {Barmby}, {Baumbach}, {Berry}, {Biscani}, {Boquien}, {Bostroem},
  {Bouma}, {Brammer}, {Bray}, {Breytenbach}, {Buddelmeijer}, {Burke},
  {Calderone}, {Cano Rodr{\'\i}guez}, {Cara}, {Cardoso}, {Cheedella}, {Copin},
  {Corrales}, {Crichton}, {D'Avella}, {Deil}, {Depagne}, {Dietrich}, {Donath},
  {Droettboom}, {Earl}, {Erben}, {Fabbro}, {Ferreira}, {Finethy}, {Fox},
  {Garrison}, {Gibbons}, {Goldstein}, {Gommers}, {Greco}, {Greenfield},
  {Groener}, {Grollier}, {Hagen}, {Hirst}, {Homeier}, {Horton}, {Hosseinzadeh},
  {Hu}, {Hunkeler}, {Ivezi{\'c}}, {Jain}, {Jenness}, {Kanarek}, {Kendrew},
  {Kern}, {Kerzendorf}, {Khvalko}, {King}, {Kirkby}, {Kulkarni}, {Kumar},
  {Lee}, {Lenz}, {Littlefair}, {Ma}, {Macleod}, {Mastropietro}, {McCully},
  {Montagnac}, {Morris}, {Mueller}, {Mumford}, {Muna}, {Murphy}, {Nelson},
  {Nguyen}, {Ninan}, {N{\"o}the}, {Ogaz}, {Oh}, {Parejko}, {Parley}, {Pascual},
  {Patil}, {Patil}, {Plunkett}, {Prochaska}, {Rastogi}, {Reddy Janga},
  {Sabater}, {Sakurikar}, {Seifert}, {Sherbert}, {Sherwood-Taylor}, {Shih},
  {Sick}, {Silbiger}, {Singanamalla}, {Singer}, {Sladen}, {Sooley},
  {Sornarajah}, {Streicher}, {Teuben}, {Thomas}, {Tremblay}, {Turner},
  {Terr{\'o}n}, {van Kerkwijk}, {de la Vega}, {Watkins}, {Weaver}, {Whitmore},
  {Woillez}, {Zabalza}, \& {Astropy Contributors}}]{astropy2}
{Astropy Collaboration}, {Price-Whelan}, A.~M., {Sip{\H{o}}cz}, B.~M., {et~al.}
  2018, \aj, 156, 123

\bibitem[{{Astropy Collaboration} {et~al.}(2022){Astropy Collaboration},
  {Price-Whelan}, {Lim}, {Earl}, {Starkman}, {Bradley}, {Shupe}, {Patil},
  {Corrales}, {Brasseur}, {N{\"o}the}, {Donath}, {Tollerud}, {Morris},
  {Ginsburg}, {Vaher}, {Weaver}, {Tocknell}, {Jamieson}, {van Kerkwijk},
  {Robitaille}, {Merry}, {Bachetti}, {G{\"u}nther}, {Aldcroft},
  {Alvarado-Montes}, {Archibald}, {B{\'o}di}, {Bapat}, {Barentsen},
  {Baz{\'a}n}, {Biswas}, {Boquien}, {Burke}, {Cara}, {Cara}, {Conroy},
  {Conseil}, {Craig}, {Cross}, {Cruz}, {D'Eugenio}, {Dencheva}, {Devillepoix},
  {Dietrich}, {Eigenbrot}, {Erben}, {Ferreira}, {Foreman-Mackey}, {Fox},
  {Freij}, {Garg}, {Geda}, {Glattly}, {Gondhalekar}, {Gordon}, {Grant},
  {Greenfield}, {Groener}, {Guest}, {Gurovich}, {Handberg}, {Hart},
  {Hatfield-Dodds}, {Homeier}, {Hosseinzadeh}, {Jenness}, {Jones}, {Joseph},
  {Kalmbach}, {Karamehmetoglu}, {Ka{\l}uszy{\'n}ski}, {Kelley}, {Kern},
  {Kerzendorf}, {Koch}, {Kulumani}, {Lee}, {Ly}, {Ma}, {MacBride}, {Maljaars},
  {Muna}, {Murphy}, {Norman}, {O'Steen}, {Oman}, {Pacifici}, {Pascual},
  {Pascual-Granado}, {Patil}, {Perren}, {Pickering}, {Rastogi}, {Roulston},
  {Ryan}, {Rykoff}, {Sabater}, {Sakurikar}, {Salgado}, {Sanghi}, {Saunders},
  {Savchenko}, {Schwardt}, {Seifert-Eckert}, {Shih}, {Jain}, {Shukla}, {Sick},
  {Simpson}, {Singanamalla}, {Singer}, {Singhal}, {Sinha}, {Sip{\H{o}}cz},
  {Spitler}, {Stansby}, {Streicher}, {{\v{S}}umak}, {Swinbank}, {Taranu},
  {Tewary}, {Tremblay}, {de Val-Borro}, {Van Kooten}, {Vasovi{\'c}}, {Verma},
  {de Miranda Cardoso}, {Williams}, {Wilson}, {Winkel}, {Wood-Vasey}, {Xue},
  {Yoachim}, {Zhang}, {Zonca}, \& {Astropy Project Contributors}}]{astropy3}
{Astropy Collaboration}, {Price-Whelan}, A.~M., {Lim}, P.~L., {et~al.} 2022,
  \apj, 935, 167

\bibitem[{{Bailer-Jones} {et~al.}(2021){Bailer-Jones}, {Rybizki}, {Fouesneau},
  {Demleitner}, \& {Andrae}}]{bailer-jones++2021-edr3-distances}
{Bailer-Jones}, C.~A.~L., {Rybizki}, J., {Fouesneau}, M., {Demleitner}, M., \&
  {Andrae}, R. 2021, \aj, 161, 147

\bibitem[{{Ball} \&
  {Gizon}(2014)}]{ball+2014-surface-correction-inertia-weighted}
{Ball}, W.~H., \& {Gizon}, L. 2014, \aap, 568, A123

\bibitem[{{Ball} {et~al.}(2018){Ball}, {Theme{\ss}l}, \&
  {Hekker}}]{ball++2018-surface-effect-rgb}
{Ball}, W.~H., {Theme{\ss}l}, N., \& {Hekker}, S. 2018, \mnras, 478, 4697

\bibitem[{{Basu} \& {Antia}(2004)}]{basu+2004-solar-helium}
{Basu}, S., \& {Antia}, H.~M. 2004, \apjl, 606, L85

\bibitem[{{Basu} \& {Chaplin}(2017)}]{basu+2017-book}
{Basu}, S., \& {Chaplin}, W.~J. 2017, {Asteroseismic Data Analysis: Foundations
  and Techniques}

\bibitem[{{Belkacem} {et~al.}(2011){Belkacem}, {Goupil}, {Dupret}, {Samadi},
  {Baudin}, {Noels}, \& {Mosser}}]{belkacem++2011-physics-under-numax-nuc}
{Belkacem}, K., {Goupil}, M.~J., {Dupret}, M.~A., {et~al.} 2011, \aap, 530,
  A142

\bibitem[{{Bellinger} {et~al.}(2019){Bellinger}, {Hekker}, {Angelou},
  {Stokholm}, \& {Basu}}]{bellinger++2019-errors}
{Bellinger}, E.~P., {Hekker}, S., {Angelou}, G.~C., {Stokholm}, A., \& {Basu},
  S. 2019, \aap, 622, A130

\bibitem[{{Benbakoura} {et~al.}(2021){Benbakoura}, {Gaulme}, {McKeever},
  {Sekaran}, {Beck}, {Spada}, {Jackiewicz}, {Mathis}, {Mathur}, {Tkachenko}, \&
  {Garc{\'\i}a}}]{benbakoura++2021-binary}
{Benbakoura}, M., {Gaulme}, P., {McKeever}, J., {et~al.} 2021, arXiv e-prints,
  arXiv:2101.05351

\bibitem[{{Berger} {et~al.}(2020){Berger}, {Huber}, {van Saders}, {Gaidos},
  {Tayar}, \& {Kraus}}]{berger++2020-gaia-kepler-1-stars}
{Berger}, T.~A., {Huber}, D., {van Saders}, J.~L., {et~al.} 2020, \aj, 159, 280

\bibitem[{{Brogaard} {et~al.}(2018){Brogaard}, {Hansen}, {Miglio}, {Slumstrup},
  {Frandsen}, {Jessen-Hansen}, {Lund}, {Bossini}, {Thygesen}, {Davies},
  {Chaplin}, {Arentoft}, {Bruntt}, {Grundahl}, \&
  {Handberg}}]{brogaard++2018-accuracy-scaling-relation}
{Brogaard}, K., {Hansen}, C.~J., {Miglio}, A., {et~al.} 2018, \mnras, 476, 3729

\bibitem[{{Broomhall} {et~al.}(2014){Broomhall}, {Miglio}, {Montalb{\'a}n},
  {Eggenberger}, {Chaplin}, {Elsworth}, {Scuflaire}, {Ventura}, \&
  {Verner}}]{broomhall++2014-helium-envelope-rg-kepler}
{Broomhall}, A.~M., {Miglio}, A., {Montalb{\'a}n}, J., {et~al.} 2014, \mnras,
  440, 1828

\bibitem[{{Brown} {et~al.}(1991){Brown}, {Gilliland}, {Noyes}, \&
  {Ramsey}}]{brown++1991-dection-procyon-scaling-relation}
{Brown}, T.~M., {Gilliland}, R.~L., {Noyes}, R.~W., \& {Ramsey}, L.~W. 1991,
  \apj, 368, 599

\bibitem[{{Casagrande} {et~al.}(2007){Casagrande}, {Flynn}, {Portinari},
  {Girardi}, \& {Jimenez}}]{casagrande++2007-He}
{Casagrande}, L., {Flynn}, C., {Portinari}, L., {Girardi}, L., \& {Jimenez}, R.
  2007, \mnras, 382, 1516

\bibitem[{{Casagrande} {et~al.}(2021){Casagrande}, {Lin}, {Rains}, {Liu},
  {Buder}, {Horner}, {Asplund}, {Lewis}, {Martell}, {Nordlander}, {Stello},
  {Ting}, {Wittenmyer}, {Bland-Hawthorn}, {Casey}, {De Silva}, {D'Orazi},
  {Freeman}, {Hayden}, {Kos}, {Lind}, {Schlesinger}, {Sharma}, {Simpson},
  {Zucker}, \& {Zwitter}}]{casagrande++2021-irfm-gaia}
{Casagrande}, L., {Lin}, J., {Rains}, A.~D., {et~al.} 2021, \mnras, 507, 2684

\bibitem[{{Chaplin} \& {Miglio}(2013)}]{chaplin+2013-solar-like-review}
{Chaplin}, W.~J., \& {Miglio}, A. 2013, \araa, 51, 353

\bibitem[{{Chen} {et~al.}(2014){Chen}, {Girardi}, {Bressan}, {Marigo},
  {Barbieri}, \& {Kong}}]{cheny++2014-parsec}
{Chen}, Y., {Girardi}, L., {Bressan}, A., {et~al.} 2014, \mnras, 444, 2525

\bibitem[{{Choi} {et~al.}(2016){Choi}, {Dotter}, {Conroy}, {Cantiello},
  {Paxton}, \& {Johnson}}]{Choi++2016-mist-1-solar-scaled-models}
{Choi}, J., {Dotter}, A., {Conroy}, C., {et~al.} 2016, \apj, 823, 102

\bibitem[{{Choi} {et~al.}(2018){Choi}, {Dotter}, {Conroy}, \&
  {Ting}}]{Choi++2018-rgb-boundary-condition-100K-teff-uncertain}
{Choi}, J., {Dotter}, A., {Conroy}, C., \& {Ting}, Y.-S. 2018, \apj, 860, 131

\bibitem[{{Chontos} {et~al.}(2021){Chontos}, {Huber}, {Sayeed}, \&
  {Yamsiri}}]{chontos++2021-pysyd}
{Chontos}, A., {Huber}, D., {Sayeed}, M., \& {Yamsiri}, P. 2021, arXiv
  e-prints, arXiv:2108.00582

\bibitem[{{Christensen-Dalsgaard}(1984)}]{jcd-1984-review}
{Christensen-Dalsgaard}, J. 1984, in Space Research in Stellar Activity and
  Variability, ed. A.~{Mangeney} \& F.~{Praderie}, 11

\bibitem[{{Christensen-Dalsgaard}(2015)}]{jcd-2015-rgb-bump}
{Christensen-Dalsgaard}, J. 2015, \mnras, 453, 666

\bibitem[{{Cinquegrana} {et~al.}(2023){Cinquegrana}, {Joyce}, \&
  {Karakas}}]{cinquegrana++2023-bridge2}
{Cinquegrana}, G.~C., {Joyce}, M., \& {Karakas}, A.~I. 2023, \mnras, 525, 3216

\bibitem[{{Claret} \& {Torres}(2018)}]{claret++2018-eb-core-fov}
{Claret}, A., \& {Torres}, G. 2018, \apj, 859, 100

\bibitem[{{Coelho} {et~al.}(2015){Coelho}, {Chaplin}, {Basu}, {Serenelli},
  {Miglio}, \& {Reese}}]{coelho++2015-numax}
{Coelho}, H.~R., {Chaplin}, W.~J., {Basu}, S., {et~al.} 2015, \mnras, 451, 3011

\bibitem[{{Creevey} {et~al.}(2017){Creevey}, {Metcalfe}, {Schultheis},
  {Salabert}, {Bazot}, {Th{\'e}venin}, {Mathur}, {Xu}, \&
  {Garc{\'\i}a}}]{creevey++2017-amp}
{Creevey}, O.~L., {Metcalfe}, T.~S., {Schultheis}, M., {et~al.} 2017, \aap,
  601, A67

\bibitem[{{Cunha} {et~al.}(2021){Cunha}, {Roxburgh}, {Aguirre B{\o}rsen-Koch},
  {Ball}, {Basu}, {Chaplin}, {Goupil}, {Nsamba}, {Ong}, {Reese}, {Verma},
  {Belkacem}, {Campante}, {Christensen-Dalsgaard}, {Clara}, {Deheuvels},
  {Monteiro}, {Noll}, {Ouazzani}, {R{\o}rsted}, {Stokholm}, \&
  {Winther}}]{cunha++2021-modelling-errors}
{Cunha}, M.~S., {Roxburgh}, I.~W., {Aguirre B{\o}rsen-Koch}, V., {et~al.} 2021,
  \mnras, 508, 5864

\bibitem[{{Cutri} {et~al.}(2003){Cutri}, {Skrutskie}, {van Dyk}, {Beichman},
  {Carpenter}, {Chester}, {Cambresy}, {Evans}, {Fowler}, {Gizis}, {Howard},
  {Huchra}, {Jarrett}, {Kopan}, {Kirkpatrick}, {Light}, {Marsh}, {McCallon},
  {Schneider}, {Stiening}, {Sykes}, {Weinberg}, {Wheaton}, {Wheelock}, \&
  {Zacarias}}]{cutri++2003-2mass-ps}
{Cutri}, R.~M., {Skrutskie}, M.~F., {van Dyk}, S., {et~al.} 2003, {2MASS All
  Sky Catalog of point sources.}

\bibitem[{{Davies} \& {Miglio}(2016)}]{davies++2016-age-review}
{Davies}, G.~R., \& {Miglio}, A. 2016, Astronomische Nachrichten, 337, 774

\bibitem[{{Deheuvels} {et~al.}(2016){Deheuvels}, {Brand{\~a}o}, {Silva
  Aguirre}, {Ballot}, {Michel}, {Cunha}, {Lebreton}, \&
  {Appourchaux}}]{deheuvels++2016-overshoot}
{Deheuvels}, S., {Brand{\~a}o}, I., {Silva Aguirre}, V., {et~al.} 2016, \aap,
  589, A93

\bibitem[{{Deheuvels} \& {Michel}(2010)}]{deheuvels++2010-avoided-crossing}
{Deheuvels}, S., \& {Michel}, E. 2010, \apss, 328, 259

\bibitem[{{Demarque} {et~al.}(2004){Demarque}, {Woo}, {Kim}, \&
  {Yi}}]{demarque++2004-yy}
{Demarque}, P., {Woo}, J.-H., {Kim}, Y.-C., \& {Yi}, S.~K. 2004, \apjs, 155,
  667

\bibitem[{{Dotter} {et~al.}(2008){Dotter}, {Chaboyer}, {Jevremovi{\'c}},
  {Kostov}, {Baron}, \& {Ferguson}}]{dotter++2008-dsep}
{Dotter}, A., {Chaboyer}, B., {Jevremovi{\'c}}, D., {et~al.} 2008, \apjs, 178,
  89

\bibitem[{{Dr{\'e}au} {et~al.}(2020){Dr{\'e}au}, {Cunha}, {Vrard}, \&
  {Avelino}}]{dreau++2020-helium}
{Dr{\'e}au}, G., {Cunha}, M.~S., {Vrard}, M., \& {Avelino}, P.~P. 2020, \mnras,
  497, 1008

\bibitem[{{Eddington}(1926)}]{eddington-1926-star}
{Eddington}, A.~S. 1926, {The Internal Constitution of the Stars} (The
  University Press)

\bibitem[{{Epstein} {et~al.}(2014){Epstein}, {Elsworth}, {Johnson}, {Shetrone},
  {Mosser}, {Hekker}, {Tayar}, {Harding}, {Pinsonneault}, {Silva Aguirre},
  {Basu}, {Beers}, {Bizyaev}, {Bedding}, {Chaplin}, {Frinchaboy},
  {Garc{\'\i}a}, {Garc{\'\i}a P{\'e}rez}, {Hearty}, {Huber}, {Ivans},
  {Majewski}, {Mathur}, {Nidever}, {Serenelli}, {Schiavon}, {Schneider},
  {Sch{\"o}nrich}, {Sobeck}, {Stassun}, {Stello}, \&
  {Zasowski}}]{epstein++2014-sc-metal-poor}
{Epstein}, C.~R., {Elsworth}, Y.~P., {Johnson}, J.~A., {et~al.} 2014, \apjl,
  785, L28

\bibitem[{{Feiden} \& {Chaboyer}(2014)}]{feiden++2014}
{Feiden}, G.~A., \& {Chaboyer}, B. 2014, \apj, 789, 53

\bibitem[{{Gaulme} {et~al.}(2016){Gaulme}, {McKeever}, {Jackiewicz}, {Rawls},
  {Corsaro}, {Mosser}, {Southworth}, {Mahadevan}, {Bender}, \&
  {Deshpande}}]{gaulme++2016-eb-sc}
{Gaulme}, P., {McKeever}, J., {Jackiewicz}, J., {et~al.} 2016, \apj, 832, 121

\bibitem[{{Gaulme} {et~al.}(2020){Gaulme}, {Jackiewicz}, {Spada}, {Chojnowski},
  {Mosser}, {McKeever}, {Hedlund}, {Vrard}, {Benbakoura}, \&
  {Damiani}}]{gaulme++20-active-rg}
{Gaulme}, P., {Jackiewicz}, J., {Spada}, F., {et~al.} 2020, \aap, 639, A63

\bibitem[{{Gehan} {et~al.}(2024){Gehan}, {Godoy-Rivera}, \&
  {Gaulme}}]{gehan++2024-rg-activity}
{Gehan}, C., {Godoy-Rivera}, D., \& {Gaulme}, P. 2024, \aap, 686, A93

\bibitem[{{Gent} {et~al.}(2022){Gent}, {Bergemann}, {Serenelli}, {Casagrande},
  {Gerber}, {Heiter}, {Kovalev}, {Morel}, {Nardetto}, {Adibekyan}, {Silva
  Aguirre}, {Asplund}, {Belkacem}, {del Burgo}, {Bigot}, {Chiavassa},
  {Rodr{\'\i}guez D{\'\i}az}, {Goupil}, {Gonz{\'a}lez Hern{\'a}ndez},
  {Mourard}, {Merle}, {M{\'e}sz{\'a}ros}, {Marshall}, {Ouazzani}, {Plez},
  {Reese}, {Trampedach}, \& {Tsantaki}}]{gent++2022-sapp}
{Gent}, M.~R., {Bergemann}, M., {Serenelli}, A., {et~al.} 2022, \aap, 658, A147

\bibitem[{{Goldreich} {et~al.}(1994){Goldreich}, {Murray}, \&
  {Kumar}}]{goldreich++1994-excitation-solar-pmodes}
{Goldreich}, P., {Murray}, N., \& {Kumar}, P. 1994, \apj, 424, 466

\bibitem[{{Gough}(1986)}]{gough-1986-aymptotic-relation}
{Gough}, D.~O. 1986, in Hydrodynamic and Magnetodynamic Problems in the Sun and
  Stars, ed. Y.~{Osaki}, 117

\bibitem[{{Green} {et~al.}(2019){Green}, {Schlafly}, {Zucker}, {Speagle}, \&
  {Finkbeiner}}]{green++2019-dustmap}
{Green}, G.~M., {Schlafly}, E., {Zucker}, C., {Speagle}, J.~S., \&
  {Finkbeiner}, D. 2019, \apj, 887, 93

\bibitem[{{Grevesse} \& {Sauval}(1998)}]{grevesse+1998-abundance}
{Grevesse}, N., \& {Sauval}, A.~J. 1998, \ssr, 85, 161

\bibitem[{{Hall} {et~al.}(2019){Hall}, {Davies}, {Elsworth}, {Miglio},
  {Bedding}, {Brown}, {Khan}, {Hawkins}, {Garc{\'\i}a}, {Chaplin}, \&
  {North}}]{hall++2019-rc-gaiadr2-seismo}
{Hall}, O.~J., {Davies}, G.~R., {Elsworth}, Y.~P., {et~al.} 2019, \mnras, 486,
  3569

\bibitem[{{Hekker}(2020)}]{hekker-2020-scaling-review}
{Hekker}, S. 2020, Frontiers in Astronomy and Space Sciences, 7, 3

\bibitem[{{Henyey} {et~al.}(1965){Henyey}, {Vardya}, \&
  {Bodenheimer}}]{henyey++1965-mlt}
{Henyey}, L., {Vardya}, M.~S., \& {Bodenheimer}, P. 1965, \apj, 142, 841

\bibitem[{{Herwig}(2000)}]{herwig-2000-overshoot}
{Herwig}, F. 2000, \aap, 360, 952

\bibitem[{{Ho} {et~al.}(2017){Ho}, {Rix}, {Ness}, {Hogg}, {Liu}, \&
  {Ting}}]{ho++2017-lamost}
{Ho}, A. Y.~Q., {Rix}, H.-W., {Ness}, M.~K., {et~al.} 2017, \apj, 841, 40

\bibitem[{{Howe} {et~al.}(2020){Howe}, {Chaplin}, {Basu}, {Ball}, {Davies},
  {Elsworth}, {Hale}, {Miglio}, {Nielsen}, \& {Viani}}]{howe++2020-numax-solar}
{Howe}, R., {Chaplin}, W.~J., {Basu}, S., {et~al.} 2020, \mnras, 493, L49

\bibitem[{{Huber} {et~al.}(2009){Huber}, {Stello}, {Bedding}, {Chaplin},
  {Arentoft}, {Quirion}, \& {Kjeldsen}}]{huber++2009-syd-pipeline}
{Huber}, D., {Stello}, D., {Bedding}, T.~R., {et~al.} 2009, Communications in
  Asteroseismology, 160, 74

\bibitem[{{Huber} {et~al.}(2017){Huber}, {Zinn}, {Bojsen-Hansen},
  {Pinsonneault}, {Sahlholdt}, {Serenelli}, {Silva Aguirre}, {Stassun},
  {Stello}, {Tayar}, {Bastien}, {Bedding}, {Buchhave}, {Chaplin}, {Davies},
  {Garc{\'\i}a}, {Latham}, {Mathur}, {Mosser}, \&
  {Sharma}}]{huber++2017-seismic-radii-gaia}
{Huber}, D., {Zinn}, J., {Bojsen-Hansen}, M., {et~al.} 2017, \apj, 844, 102

\bibitem[{Hunter(2007)}]{matplotlib}
Hunter, J.~D. 2007, Computing in Science \& Engineering, 9, 90

\bibitem[{{Jackson} \& {Jeffries}(2014)}]{jackson++2014}
{Jackson}, R.~J., \& {Jeffries}, R.~D. 2014, \mnras, 441, 2111

\bibitem[{{Jermyn} {et~al.}(2023){Jermyn}, {Bauer}, {Schwab}, {Farmer}, {Ball},
  {Bellinger}, {Dotter}, {Joyce}, {Marchant}, {Mombarg}, {Wolf}, {Sunny Wong},
  {Cinquegrana}, {Farrell}, {Smolec}, {Thoul}, {Cantiello}, {Herwig}, {Toloza},
  {Bildsten}, {Townsend}, \& {Timmes}}]{jermyn++2023-mesa}
{Jermyn}, A.~S., {Bauer}, E.~B., {Schwab}, J., {et~al.} 2023, \apjs, 265, 15

\bibitem[{{Jiang} {et~al.}(2023){Jiang}, {Wu}, {Feinstein}, {Stassun},
  {Bedding}, {Veras}, {Corsaro}, {Buzasi}, {Stello}, {Li}, {Mathur},
  {Garc{\'\i}a}, {Breton}, {Lundkvist}, {Miko{\l}ajczyk}, {Gehan}, {Campante},
  {Bossini}, {Kane}, {Joel Ong}, {Y{\i}ld{\i}z}, {Kayhan}, {{\c{C}}elik Orhan},
  {{\"O}rtel}, {Zhang}, {Cunha}, {de Moura}, {Yu}, {Huber}, {Ou}, {Wittenmyer},
  {Gizon}, \& {Chaplin}}]{jiangc++2023-hd76920}
{Jiang}, C., {Wu}, T., {Feinstein}, A.~D., {et~al.} 2023, \apj, 945, 20

\bibitem[{{Jim{\'e}nez} {et~al.}(2015){Jim{\'e}nez}, {Garc{\'\i}a}, {P{\'e}rez
  Hern{\'a}ndez}, \& {Mathur}}]{jimenez++2015-nuac-six-kepler}
{Jim{\'e}nez}, A., {Garc{\'\i}a}, R.~A., {P{\'e}rez Hern{\'a}ndez}, F., \&
  {Mathur}, S. 2015, \aap, 583, A74

\bibitem[{{Jimenez} {et~al.}(2003){Jimenez}, {Flynn}, {MacDonald}, \&
  {Gibson}}]{jimenez++2003-He}
{Jimenez}, R., {Flynn}, C., {MacDonald}, J., \& {Gibson}, B.~K. 2003, Science,
  299, 1552

\bibitem[{{Joyce} \& {Chaboyer}(2018{\natexlab{a}})}]{joyce+2018-alpha-cen}
{Joyce}, M., \& {Chaboyer}, B. 2018{\natexlab{a}}, \apj, 864, 99

\bibitem[{{Joyce} \&
  {Chaboyer}(2018{\natexlab{b}})}]{joyce+2018-amlt-metal-poor}
---. 2018{\natexlab{b}}, \apj, 856, 10

\bibitem[{{Joyce} {et~al.}(2023){Joyce}, {Johnson}, {Marchetti}, {Rich},
  {Simion}, \& {Bourke}}]{joyce++2023-age}
{Joyce}, M., {Johnson}, C.~I., {Marchetti}, T., {et~al.} 2023, \apj, 946, 28

\bibitem[{{Joyce} \& {Tayar}(2023)}]{joyce+2023-mlt}
{Joyce}, M., \& {Tayar}, J. 2023, Galaxies, 11, 75

\bibitem[{{Kallinger}(2019)}]{kallinger++2019-peakbagging}
{Kallinger}, T. 2019, arXiv e-prints, arXiv:1906.09428

\bibitem[{{Kallinger} {et~al.}(2018){Kallinger}, {Beck}, {Stello}, \&
  {Garcia}}]{kallinger++2018-sc}
{Kallinger}, T., {Beck}, P.~G., {Stello}, D., \& {Garcia}, R.~A. 2018, \aap,
  616, A104

\bibitem[{{Kallinger} {et~al.}(2012){Kallinger}, {Hekker}, {Mosser}, {De
  Ridder}, {Bedding}, {Elsworth}, {Gruberbauer}, {Guenther}, {Stello}, {Basu},
  {Garc{\'\i}a}, {Chaplin}, {Mullally}, {Still}, \&
  {Thompson}}]{kallinger++2012-epsp}
{Kallinger}, T., {Hekker}, S., {Mosser}, B., {et~al.} 2012, \aap, 541, A51

\bibitem[{{Karlsmose}(2019)}]{Karlsmose-thesis}
{Karlsmose}, K.~G. 2019, Master's thesis, Aarhus University

\bibitem[{{Khan} {et~al.}(2018){Khan}, {Hall}, {Miglio}, {Davies}, {Mosser},
  {Girardi}, \&
  {Montalb{\'a}n}}]{khan-2018-rgb-bump-constraints-envelope-overshooting}
{Khan}, S., {Hall}, O.~J., {Miglio}, A., {et~al.} 2018, \apj, 859, 156

\bibitem[{{Khan} {et~al.}(2019){Khan}, {Miglio}, {Mosser}, {Arenou},
  {Belkacem}, {Brown}, {Katz}, {Casagrand e}, {Chaplin}, {Davies}, {Rendle},
  {Rodrigues}, {Bossini}, {Cantat-Gaudin}, {Elsworth}, {Girardi}, {North}, \&
  {Vallenari}}]{khan++2019-gaiadr2-zero-point}
{Khan}, S., {Miglio}, A., {Mosser}, B., {et~al.} 2019, \aap, 628, A35

\bibitem[{{Kjeldsen} \& {Bedding}(1995)}]{kjeldsen+1995-scaling-relations}
{Kjeldsen}, H., \& {Bedding}, T.~R. 1995, \aap, 293, 87

\bibitem[{{Lebreton} \& {Goupil}(2014)}]{lebreton+2014-hd52265}
{Lebreton}, Y., \& {Goupil}, M.~J. 2014, \aap, 569, A21

\bibitem[{{Lebreton} {et~al.}(2014){Lebreton}, {Goupil}, \&
  {Montalb{\'a}n}}]{lebreton++2014-age}
{Lebreton}, Y., {Goupil}, M.~J., \& {Montalb{\'a}n}, J. 2014, in EAS
  Publications Series, Vol.~65, EAS Publications Series, ed. Y.~{Lebreton},
  D.~{Valls-Gabaud}, \& C.~{Charbonnel}, 99--176

\bibitem[{{Li} {et~al.}(2020){Li}, {Bedding}, {Christensen-Dalsgaard},
  {Stello}, {Li}, \& {Keen}}]{litd++2020-kepler-36-subgiants}
{Li}, T., {Bedding}, T.~R., {Christensen-Dalsgaard}, J., {et~al.} 2020, \mnras,
  495, 3431

\bibitem[{{Li} {et~al.}(2018){Li}, {Bedding}, {Huber}, {Ball}, {Stello},
  {Murphy}, \& {Bland -Hawthorn}}]{litd++2018-amlt}
{Li}, T., {Bedding}, T.~R., {Huber}, D., {et~al.} 2018, \mnras, 475, 981

\bibitem[{{Li} {et~al.}(2022){Li}, {Li}, {Bi}, {Bedding}, {Davies}, \&
  {Du}}]{litd++2022-kepler-rgb}
{Li}, T., {Li}, Y., {Bi}, S., {et~al.} 2022, \apj, 927, 167

\bibitem[{{Li} {et~al.}(2021){Li}, {Bedding}, {Stello}, {Sharma}, {Huber}, \&
  {Murphy}}]{liyg++2021-sc-intrinsic-scatter}
{Li}, Y., {Bedding}, T.~R., {Stello}, D., {et~al.} 2021, \mnras, 501, 3162

\bibitem[{{Li} {et~al.}(2023){Li}, {Bedding}, {Stello}, {Huber}, {Hon},
  {Joyce}, {Li}, {Perkins}, {White}, {Zinn}, {Howard}, {Isaacson}, {Hey}, \&
  {Kjeldsen}}]{liyg++2022-surface}
---. 2023, \mnras, 523, 916

\bibitem[{{Lightkurve Collaboration} {et~al.}(2018){Lightkurve Collaboration},
  {Cardoso}, {Hedges}, {Gully-Santiago}, {Saunders}, {Cody}, {Barclay}, {Hall},
  {Sagear}, {Turtelboom}, {Zhang}, {Tzanidakis}, {Mighell}, {Coughlin}, {Bell},
  {Berta-Thompson}, {Williams}, {Dotson}, \& {Barentsen}}]{lightkurve}
{Lightkurve Collaboration}, {Cardoso}, J.~V.~d.~M., {Hedges}, C., {et~al.}
  2018, {Lightkurve: Kepler and TESS time series analysis in Python},
  Astrophysics Source Code Library

\bibitem[{{Lindsay} {et~al.}(2024){Lindsay}, {Ong}, \&
  {Basu}}]{lindsay++2024-ov}
{Lindsay}, C.~J., {Ong}, J.~M.~J., \& {Basu}, S. 2024, \apj, 965, 171

\bibitem[{{Magic} {et~al.}(2013){Magic}, {Collet}, {Asplund}, {Trampedach},
  {Hayek}, {Chiavassa}, {Stein}, \& {Nordlund}}]{magic++2013-stagger-1}
{Magic}, Z., {Collet}, R., {Asplund}, M., {et~al.} 2013, \aap, 557, A26

\bibitem[{{Magic} {et~al.}(2015){Magic}, {Weiss}, \&
  {Asplund}}]{magic++2015-stagger-3-amlt}
{Magic}, Z., {Weiss}, A., \& {Asplund}, M. 2015, \aap, 573, A89

\bibitem[{{McKeever} {et~al.}(2019){McKeever}, {Basu}, \&
  {Corsaro}}]{mckkever++2019-ngc6791}
{McKeever}, J.~M., {Basu}, S., \& {Corsaro}, E. 2019, \apj, 874, 180

\bibitem[{{Metcalfe} {et~al.}(2014){Metcalfe}, {Creevey}, {Do{\u{g}}an},
  {Mathur}, {Xu}, {Bedding}, {Chaplin}, {Christensen-Dalsgaard}, {Karoff},
  {Trampedach}, {Benomar}, {Brown}, {Buzasi}, {Campante}, {{\c{C}}elik},
  {Cunha}, {Davies}, {Deheuvels}, {Derekas}, {Di Mauro}, {Garc{\'\i}a},
  {Guzik}, {Howe}, {MacGregor}, {Mazumdar}, {Montalb{\'a}n}, {Monteiro},
  {Salabert}, {Serenelli}, {Stello}, {Ste\&{\c{s}}acute}, {licki}, {Suran},
  {Y{\i}ld{\i}z}, {Aksoy}, {Elsworth}, {Gruberbauer}, {Guenther}, {Lebreton},
  {Molaverdikhani}, {Pricopi}, {Simoniello}, \& {White}}]{metcalfe++2014-amp}
{Metcalfe}, T.~S., {Creevey}, O.~L., {Do{\u{g}}an}, G., {et~al.} 2014, \apjs,
  214, 27

\bibitem[{{Miglio} {et~al.}(2021){Miglio}, {Chiappini}, {Mackereth}, {Davies},
  {Brogaard}, {Casagrande}, {Chaplin}, {Girardi}, {Kawata}, {Khan}, {Izzard},
  {Montalb{\'a}n}, {Mosser}, {Vincenzo}, {Bossini}, {Noels}, {Rodrigues},
  {Valentini}, \& {Mandel}}]{miglio++2021-age-kepler}
{Miglio}, A., {Chiappini}, C., {Mackereth}, J.~T., {et~al.} 2021, \aap, 645,
  A85

\bibitem[{{Mombarg} {et~al.}(2021){Mombarg}, {Van Reeth}, \&
  {Aerts}}]{mombarg++2021}
{Mombarg}, J.~S.~G., {Van Reeth}, T., \& {Aerts}, C. 2021, \aap, 650, A58

\bibitem[{{Montalb{\'a}n} {et~al.}(2010){Montalb{\'a}n}, {Miglio}, {Noels},
  {Scuflaire}, \& {Ventura}}]{montalban++2010-models}
{Montalb{\'a}n}, J., {Miglio}, A., {Noels}, A., {Scuflaire}, R., \& {Ventura},
  P. 2010, \apjl, 721, L182

\bibitem[{{Montalb{\'a}n} {et~al.}(2021){Montalb{\'a}n}, {Mackereth}, {Miglio},
  {Vincenzo}, {Chiappini}, {Buldgen}, {Mosser}, {Noels}, {Scuflaire}, {Vrard},
  {Willett}, {Davies}, {Hall}, {Nielsen}, {Khan}, {Rendle}, {van Rossem},
  {Ferguson}, \& {Chaplin}}]{montalban++2021-age}
{Montalb{\'a}n}, J., {Mackereth}, J.~T., {Miglio}, A., {et~al.} 2021, Nature
  Astronomy, 5, 640

\bibitem[{{Mosser} {et~al.}(2011){Mosser}, {Belkacem}, {Goupil}, {Michel},
  {Elsworth}, {Barban}, {Kallinger}, {Hekker}, {De Ridder}, {Samadi}, {Baudin},
  {Pinheiro}, {Auvergne}, {Baglin}, \& {Catala}}]{mosser++2011-uni-pattern}
{Mosser}, B., {Belkacem}, K., {Goupil}, M.~J., {et~al.} 2011, \aap, 525, L9

\bibitem[{{Nguyen} {et~al.}(2022){Nguyen}, {Costa}, {Girardi}, {Volpato},
  {Bressan}, {Chen}, {Marigo}, {Fu}, \& {Goudfrooij}}]{nguyen++2022-parsec2}
{Nguyen}, C.~T., {Costa}, G., {Girardi}, L., {et~al.} 2022, \aap, 665, A126

\bibitem[{{Noll} {et~al.}(2024){Noll}, {Basu}, \& {Hekker}}]{noll++2024-rc}
{Noll}, A., {Basu}, S., \& {Hekker}, S. 2024, \aap, 683, A189

\bibitem[{{Ong} \& {Basu}(2020)}]{ong++2020-coupling}
{Ong}, J.~M.~J., \& {Basu}, S. 2020, \apj, 898, 127

\bibitem[{{Ong} {et~al.}(2021{\natexlab{a}}){Ong}, {Basu}, {Lund}, {Bieryla},
  {Viani}, \& {Latham}}]{ong++2021-surf-corr-sg}
{Ong}, J.~M.~J., {Basu}, S., {Lund}, M.~N., {et~al.} 2021{\natexlab{a}}, \apj,
  922, 18

\bibitem[{{Ong} {et~al.}(2021{\natexlab{b}}){Ong}, {Basu}, \&
  {Roxburgh}}]{ong++2021-surf-corr-theory}
{Ong}, J.~M.~J., {Basu}, S., \& {Roxburgh}, I.~W. 2021{\natexlab{b}}, \apj,
  920, 8

\bibitem[{{Ong} {et~al.}(2022){Ong}, {Lund}, {Basu}, {Stassun}, {Tayar},
  {Y{\i}ld{\i}z}, {{\c{C}}elik Orhan}, {{\"O}rtel}, {Antia}, {Appourchaux},
  {Corsaro}, {Davies}, {Theme{\ss}l}, {Viani}, \& {Huber}}]{ong++2022-helium}
{Ong}, J.~M.~J., {Lund}, M.~N., {Basu}, S., {et~al.} 2022, in Cambridge
  Workshop on Cool Stars, Stellar Systems, and the Sun, Cambridge Workshop on
  Cool Stars, Stellar Systems, and the Sun, 1

\bibitem[{{OpenAI}(2023)}]{gpt4}
{OpenAI}. 2023, arXiv e-prints, arXiv:2303.08774

\bibitem[{{Paxton} {et~al.}(2011){Paxton}, {Bildsten}, {Dotter}, {Herwig},
  {Lesaffre}, \& {Timmes}}]{paxton++2011-mesa}
{Paxton}, B., {Bildsten}, L., {Dotter}, A., {et~al.} 2011, \apjs, 192, 3

\bibitem[{{Paxton} {et~al.}(2013){Paxton}, {Cantiello}, {Arras}, {Bildsten},
  {Brown}, {Dotter}, {Mankovich}, {Montgomery}, {Stello}, {Timmes}, \&
  {Townsend}}]{paxton++2013-mesa}
{Paxton}, B., {Cantiello}, M., {Arras}, P., {et~al.} 2013, \apjs, 208, 4

\bibitem[{{Paxton} {et~al.}(2015){Paxton}, {Marchant}, {Schwab}, {Bauer},
  {Bildsten}, {Cantiello}, {Dessart}, {Farmer}, {Hu}, {Langer}, {Townsend},
  {Townsley}, \& {Timmes}}]{paxton++2015-mesa}
{Paxton}, B., {Marchant}, P., {Schwab}, J., {et~al.} 2015, \apjs, 220, 15

\bibitem[{{Paxton} {et~al.}(2018){Paxton}, {Schwab}, {Bauer}, {Bildsten},
  {Blinnikov}, {Duffell}, {Farmer}, {Goldberg}, {Marchant}, {Sorokina},
  {Thoul}, {Townsend}, \& {Timmes}}]{paxton++2018-mesa}
{Paxton}, B., {Schwab}, J., {Bauer}, E.~B., {et~al.} 2018, \apjs, 234, 34

\bibitem[{{Paxton} {et~al.}(2019){Paxton}, {Smolec}, {Schwab}, {Gautschy},
  {Bildsten}, {Cantiello}, {Dotter}, {Farmer}, {Goldberg}, {Jermyn}, {Kanbur},
  {Marchant}, {Thoul}, {Townsend}, {Wolf}, {Zhang}, \&
  {Timmes}}]{paxton++2019-mesa}
{Paxton}, B., {Smolec}, R., {Schwab}, J., {et~al.} 2019, \apjs, 243, 10

\bibitem[{{Peimbert} \& {Torres-Peimbert}(1974)}]{peimbert++1974}
{Peimbert}, M., \& {Torres-Peimbert}, S. 1974, \apj, 193, 327

\bibitem[{{Pinsonneault} {et~al.}(2018){Pinsonneault}, {Elsworth}, {Tayar},
  {Serenelli}, {Stello}, {Zinn}, {Mathur}, {Garc{\'\i}a}, {Johnson}, {Hekker},
  {Huber}, {Kallinger}, {M{\'e}sz{\'a}ros}, {Mosser}, {Stassun}, {Girardi},
  {Rodrigues}, {Silva Aguirre}, {An}, {Basu}, {Chaplin}, {Corsaro}, {Cunha},
  {Garc{\'\i}a-Hern{\'a}ndez}, {Holtzman}, {J{\"o}nsson}, {Shetrone}, {Smith},
  {Sobeck}, {Stringfellow}, {Zamora}, {Beers}, {Fern{\'a}ndez-Trincado},
  {Frinchaboy}, {Hearty}, \& {Nitschelm}}]{pinsonneault++2018-apokasc}
{Pinsonneault}, M.~H., {Elsworth}, Y.~P., {Tayar}, J., {et~al.} 2018, \apjs,
  239, 32

\bibitem[{{Planck Collaboration} {et~al.}(2016){Planck Collaboration}, {Ade},
  {Aghanim}, {Arnaud}, {Ashdown}, {Aumont}, {Baccigalupi}, {Banday},
  {Barreiro}, {Bartlett}, {Bartolo}, {Battaner}, {Battye}, {Benabed},
  {Beno{\^\i}t}, {Benoit-L{\'e}vy}, {Bernard}, {Bersanelli}, {Bielewicz},
  {Bock}, {Bonaldi}, {Bonavera}, {Bond}, {Borrill}, {Bouchet}, {Boulanger},
  {Bucher}, {Burigana}, {Butler}, {Calabrese}, {Cardoso}, {Catalano},
  {Challinor}, {Chamballu}, {Chary}, {Chiang}, {Chluba}, {Christensen},
  {Church}, {Clements}, {Colombi}, {Colombo}, {Combet}, {Coulais}, {Crill},
  {Curto}, {Cuttaia}, {Danese}, {Davies}, {Davis}, {de Bernardis}, {de Rosa},
  {de Zotti}, {Delabrouille}, {D{\'e}sert}, {Di Valentino}, {Dickinson},
  {Diego}, {Dolag}, {Dole}, {Donzelli}, {Dor{\'e}}, {Douspis}, {Ducout},
  {Dunkley}, {Dupac}, {Efstathiou}, {Elsner}, {En{\ss}lin}, {Eriksen},
  {Farhang}, {Fergusson}, {Finelli}, {Forni}, {Frailis}, {Fraisse},
  {Franceschi}, {Frejsel}, {Galeotta}, {Galli}, {Ganga}, {Gauthier}, {Gerbino},
  {Ghosh}, {Giard}, {Giraud-H{\'e}raud}, {Giusarma}, {Gjerl{\o}w},
  {Gonz{\'a}lez-Nuevo}, {G{\'o}rski}, {Gratton}, {Gregorio}, {Gruppuso},
  {Gudmundsson}, {Hamann}, {Hansen}, {Hanson}, {Harrison}, {Helou},
  {Henrot-Versill{\'e}}, {Hern{\'a}ndez-Monteagudo}, {Herranz}, {Hildebrandt},
  {Hivon}, {Hobson}, {Holmes}, {Hornstrup}, {Hovest}, {Huang}, {Huffenberger},
  {Hurier}, {Jaffe}, {Jaffe}, {Jones}, {Juvela}, {Keih{\"a}nen}, {Keskitalo},
  {Kisner}, {Kneissl}, {Knoche}, {Knox}, {Kunz}, {Kurki-Suonio}, {Lagache},
  {L{\"a}hteenm{\"a}ki}, {Lamarre}, {Lasenby}, {Lattanzi}, {Lawrence}, {Leahy},
  {Leonardi}, {Lesgourgues}, {Levrier}, {Lewis}, {Liguori}, {Lilje},
  {Linden-V{\o}rnle}, {L{\'o}pez-Caniego}, {Lubin}, {Mac{\'\i}as-P{\'e}rez},
  {Maggio}, {Maino}, {Mandolesi}, {Mangilli}, {Marchini}, {Maris}, {Martin},
  {Martinelli}, {Mart{\'\i}nez-Gonz{\'a}lez}, {Masi}, {Matarrese}, {McGehee},
  {Meinhold}, {Melchiorri}, {Melin}, {Mendes}, {Mennella}, {Migliaccio},
  {Millea}, {Mitra}, {Miville-Desch{\^e}nes}, {Moneti}, {Montier}, {Morgante},
  {Mortlock}, {Moss}, {Munshi}, {Murphy}, {Naselsky}, {Nati}, {Natoli},
  {Netterfield}, {N{\o}rgaard-Nielsen}, {Noviello}, {Novikov}, {Novikov},
  {Oxborrow}, {Paci}, {Pagano}, {Pajot}, {Paladini}, {Paoletti}, {Partridge},
  {Pasian}, {Patanchon}, {Pearson}, {Perdereau}, {Perotto}, {Perrotta},
  {Pettorino}, {Piacentini}, {Piat}, {Pierpaoli}, {Pietrobon}, {Plaszczynski},
  {Pointecouteau}, {Polenta}, {Popa}, {Pratt}, {Pr{\'e}zeau}, {Prunet},
  {Puget}, {Rachen}, {Reach}, {Rebolo}, {Reinecke}, {Remazeilles}, {Renault},
  {Renzi}, {Ristorcelli}, {Rocha}, {Rosset}, {Rossetti}, {Roudier},
  {Rouill{\'e} d'Orfeuil}, {Rowan-Robinson}, {Rubi{\~n}o-Mart{\'\i}n},
  {Rusholme}, {Said}, {Salvatelli}, {Salvati}, {Sandri}, {Santos},
  {Savelainen}, {Savini}, {Scott}, {Seiffert}, {Serra}, {Shellard}, {Spencer},
  {Spinelli}, {Stolyarov}, {Stompor}, {Sudiwala}, {Sunyaev}, {Sutton},
  {Suur-Uski}, {Sygnet}, {Tauber}, {Terenzi}, {Toffolatti}, {Tomasi},
  {Tristram}, {Trombetti}, {Tucci}, {Tuovinen}, {T{\"u}rler}, {Umana},
  {Valenziano}, {Valiviita}, {Van Tent}, {Vielva}, {Villa}, {Wade}, {Wandelt},
  {Wehus}, {White}, {White}, {Wilkinson}, {Yvon}, {Zacchei}, \&
  {Zonca}}]{planck++2015-primordial-helium}
{Planck Collaboration}, {Ade}, P.~A.~R., {Aghanim}, N., {et~al.} 2016, \aap,
  594, A13

\bibitem[{{Reyes} {et~al.}(2024){Reyes}, {Stello}, {Hon}, {Trampedach},
  {Sandquist}, \& {Pinsonneault}}]{reyes++2024-m67}
{Reyes}, C., {Stello}, D., {Hon}, M., {et~al.} 2024, arXiv e-prints,
  arXiv:2407.03526

\bibitem[{{Sahlholdt} \& {Silva
  Aguirre}(2018)}]{sahlholdt+2018-gaiadr2-sc-radius-dwarfs}
{Sahlholdt}, C.~L., \& {Silva Aguirre}, V. 2018, \mnras, 481, L125

\bibitem[{{Salaris} {et~al.}(2018){Salaris}, {Cassisi}, {Schiavon}, \&
  {Pietrinferni}}]{salaris++2018-teff-rg-apokasc-mixing-length-calibration}
{Salaris}, M., {Cassisi}, S., {Schiavon}, R.~P., \& {Pietrinferni}, A. 2018,
  \aap, 612, A68

\bibitem[{{Samadi} {et~al.}(2012){Samadi}, {Belkacem}, {Dupret}, {Ludwig},
  {Baudin}, {Caffau}, {Goupil}, \& {Barban}}]{samadi++2012-amplitudes-rg-corot}
{Samadi}, R., {Belkacem}, K., {Dupret}, M.~A., {et~al.} 2012, \aap, 543, A120

\bibitem[{{Schonhut-Stasik} {et~al.}(2023){Schonhut-Stasik}, {Zinn}, {Stassun},
  {Pinsonneault}, {Johnson}, {Warfield}, {Stello}, {Elsworth}, {Garcia},
  {Marhur}, {Mosser}, {Tayar}, {Stringfellow}, {Beaton}, {Jonsson}, \&
  {Minniti}}]{schonhut-stasik++2023-k2}
{Schonhut-Stasik}, J., {Zinn}, J.~C., {Stassun}, K.~G., {et~al.} 2023, arXiv
  e-prints, arXiv:2304.10654

\bibitem[{{Serenelli} {et~al.}(2013){Serenelli}, {Bergemann}, {Ruchti}, \&
  {Casagrande}}]{serenelli++2013-garstec}
{Serenelli}, A.~M., {Bergemann}, M., {Ruchti}, G., \& {Casagrande}, L. 2013,
  \mnras, 429, 3645

\bibitem[{{Sharma} {et~al.}(2016){Sharma}, {Stello}, {Bland-Hawthorn}, {Huber},
  \& {Bedding}}]{sharma++2016-population-rg-kepler}
{Sharma}, S., {Stello}, D., {Bland-Hawthorn}, J., {Huber}, D., \& {Bedding},
  T.~R. 2016, \apj, 822, 15

\bibitem[{{Silva Aguirre} {et~al.}(2012){Silva Aguirre}, {Casagrande}, {Basu},
  {Campante}, {Chaplin}, {Huber}, {Miglio}, {Serenelli}, {Ballot}, {Bedding},
  {Christensen-Dalsgaard}, {Creevey}, {Elsworth}, {Garc{\'\i}a}, {Gilliland},
  {Hekker}, {Kjeldsen}, {Mathur}, {Metcalfe}, {Monteiro}, {Mosser},
  {Pinsonneault}, {Stello}, {Weiss}, {Tenenbaum}, {Twicken}, \&
  {Uddin}}]{silvaaguirre++2012-seismic-parallax}
{Silva Aguirre}, V., {Casagrande}, L., {Basu}, S., {et~al.} 2012, \apj, 757, 99

\bibitem[{{Silva Aguirre} {et~al.}(2015){Silva Aguirre}, {Davies}, {Basu},
  {Christensen-Dalsgaard}, {Creevey}, {Metcalfe}, {Bedding}, {Casagrande},
  {Handberg}, {Lund}, {Nissen}, {Chaplin}, {Huber}, {Serenelli}, {Stello}, {Van
  Eylen}, {Campante}, {Elsworth}, {Gilliland}, {Hekker}, {Karoff}, {Kawaler},
  {Kjeldsen}, \& {Lundkvist}}]{silvaaguirre++2015-33-kepler-exoplanet-host}
{Silva Aguirre}, V., {Davies}, G.~R., {Basu}, S., {et~al.} 2015, \mnras, 452,
  2127

\bibitem[{{Silva Aguirre} {et~al.}(2017){Silva Aguirre}, {Lund}, {Antia},
  {Ball}, {Basu}, {Christensen-Dalsgaard}, {Lebreton}, {Reese}, {Verma},
  {Casagrande}, {Justesen}, {Mosumgaard}, {Chaplin}, {Bedding}, {Davies},
  {Handberg}, {Houdek}, {Huber}, {Kjeldsen}, {Latham}, {White}, {Coelho},
  {Miglio}, \& {Rendle}}]{silvaaguirre++2017-legacy2-modelling}
{Silva Aguirre}, V., {Lund}, M.~N., {Antia}, H.~M., {et~al.} 2017, \apj, 835,
  173

\bibitem[{{Silva Aguirre} {et~al.}(2020){Silva Aguirre}, {Stello}, {Stokholm},
  {Mosumgaard}, {Ball}, {Basu}, {Bossini}, {Bugnet}, {Buzasi}, {Campante},
  {Carboneau}, {Chaplin}, {Corsaro}, {Davies}, {Elsworth}, {Garc{\'\i}a},
  {Gaulme}, {Hall}, {Handberg}, {Hon}, {Kallinger}, {Kang}, {Lund}, {Mathur},
  {Mints}, {Mosser}, {{\c{C}}elik Orhan}, {Rodrigues}, {Vrard}, {Y{\i}ld{\i}z},
  {Zinn}, {{\"O}rtel}, {Beck}, {Bell}, {Guo}, {Jiang}, {Kuszlewicz}, {Kuehn},
  {Li}, {Lundkvist}, {Pinsonneault}, {Tayar}, {Cunha}, {Hekker}, {Huber},
  {Miglio}, {F.~G. Monteiro}, {Slumstrup}, {Winther}, {Angelou}, {Benomar},
  {B{\'o}di}, {De Moura}, {Deheuvels}, {Derekas}, {Di Mauro}, {Dupret},
  {Jim{\'e}nez}, {Lebreton}, {Matthews}, {Nardetto}, {do Nascimento},
  {Pereira}, {Rodr{\'\i}guez D{\'\i}az}, {Serenelli}, {Spitoni},
  {Stonkut{\.{e}}}, {Su{\'a}rez}, {Szab{\'o}}, {Van Eylen}, {Ventura}, {Verma},
  {Weiss}, {Wu}, {Barclay}, {Christensen-Dalsgaard}, {Jenkins}, {Kjeldsen},
  {Ricker}, {Seager}, \& {Vanderspek}}]{silvaaguirre++2020-tess-rg}
{Silva Aguirre}, V., {Stello}, D., {Stokholm}, A., {et~al.} 2020, \apjl, 889,
  L34

\bibitem[{{Smith} {et~al.}(2012){Smith}, {Stumpe}, {Van Cleve}, {Jenkins},
  {Barclay}, {Fanelli}, {Girouard}, {Kolodziejczak}, {McCauliff}, {Morris}, \&
  {Twicken}}]{smith++2012-pdc-2}
{Smith}, J.~C., {Stumpe}, M.~C., {Van Cleve}, J.~E., {et~al.} 2012, \pasp, 124,
  1000

\bibitem[{{Somers} \& {Pinsonneault}(2015)}]{somers++2015}
{Somers}, G., \& {Pinsonneault}, M.~H. 2015, \apj, 807, 174

\bibitem[{{Sreenivas} {et~al.}(2024){Sreenivas}, {Bedding}, {Li}, {Huber},
  {Stello}, {Crawford}, \& {Yu}}]{sreenivas++2024-numax}
{Sreenivas}, K.~R., {Bedding}, T.~R., {Li}, Y., {et~al.} 2024, arXiv e-prints,
  arXiv:2401.17557

\bibitem[{{Stumpe} {et~al.}(2012){Stumpe}, {Smith}, {Van Cleve}, {Twicken},
  {Barclay}, {Fanelli}, {Girouard}, {Jenkins}, {Kolodziejczak}, {McCauliff}, \&
  {Morris}}]{stumpe++2012-pdc-1}
{Stumpe}, M.~C., {Smith}, J.~C., {Van Cleve}, J.~E., {et~al.} 2012, \pasp, 124,
  985

\bibitem[{{Tassoul}(1980)}]{tassoul-1980-asymptotic-relation}
{Tassoul}, M. 1980, \apjs, 43, 469

\bibitem[{{Tayar} {et~al.}(2022){Tayar}, {Claytor}, {Huber}, \& {van
  Saders}}]{tayar++2022-error}
{Tayar}, J., {Claytor}, Z.~R., {Huber}, D., \& {van Saders}, J. 2022, \apj,
  927, 31

\bibitem[{{Tayar} {et~al.}(2017){Tayar}, {Somers}, {Pinsonneault}, {Stello},
  {Mints}, {Johnson}, {Zamora}, {Garc{\'\i}a-Hern{\'a}ndez}, {Maraston},
  {Serenelli}, {Allende Prieto}, {Bastien}, {Basu}, {Bird}, {Cohen}, {Cunha},
  {Elsworth}, {Garc{\'\i}a}, {Girardi}, {Hekker}, {Holtzman}, {Huber},
  {Mathur}, {M{\'e}sz{\'a}ros}, {Mosser}, {Shetrone}, {Silva Aguirre},
  {Stassun}, {Stringfellow}, {Zasowski}, \& {Roman-Lopes}}]{tayar-2017-amlt}
{Tayar}, J., {Somers}, G., {Pinsonneault}, M.~H., {et~al.} 2017, \apj, 840, 17

\bibitem[{{Tian} {et~al.}(2015){Tian}, {Bi}, {Bedding}, \&
  {Yang}}]{tianzj-2015-subgiant-kic6442183-kic11137075}
{Tian}, Z., {Bi}, S., {Bedding}, T.~R., \& {Yang}, W. 2015, \aap, 580, A44

\bibitem[{{Torres} \& {Ribas}(2002)}]{torres++2002-eb}
{Torres}, G., \& {Ribas}, I. 2002, \apj, 567, 1140

\bibitem[{{Townsend}(2020)}]{mesasdk}
{Townsend}, R. 2020, {MESA SDK for Linux}

\bibitem[{{Townsend} \& {Teitler}(2013)}]{townsend+2013-gyre}
{Townsend}, R.~H.~D., \& {Teitler}, S.~A. 2013, \mnras, 435, 3406

\bibitem[{{Trampedach} {et~al.}(2014){Trampedach}, {Stein},
  {Christensen-Dalsgaard}, {Nordlund}, \& {Asplund}}]{trampedach++2014-amlt}
{Trampedach}, R., {Stein}, R.~F., {Christensen-Dalsgaard}, J., {Nordlund},
  {\r{A}}., \& {Asplund}, M. 2014, \mnras, 445, 4366

\bibitem[{{Valle} {et~al.}(2019){Valle}, {Dell'Omodarme}, {Prada Moroni}, \&
  {Degl'Innocenti}}]{valle++2019-amlt}
{Valle}, G., {Dell'Omodarme}, M., {Prada Moroni}, P.~G., \& {Degl'Innocenti},
  S. 2019, \aap, 623, A59

\bibitem[{{Valle} {et~al.}(2018){Valle}, {Dell'Omodarme}, {Tognelli}, {Prada
  Moroni}, \& {Degl'Innocenti}}]{valle++2018-rgb-age}
{Valle}, G., {Dell'Omodarme}, M., {Tognelli}, E., {Prada Moroni}, P.~G., \&
  {Degl'Innocenti}, S. 2018, \aap, 619, A158

\bibitem[{{van der Walt} {et~al.}(2011){van der Walt}, {Colbert}, \&
  {Varoquaux}}]{numpy}
{van der Walt}, S., {Colbert}, S.~C., \& {Varoquaux}, G. 2011, Computing in
  Science and Engineering, 13, 22

\bibitem[{{Verma} {et~al.}(2019){Verma}, {Raodeo}, {Basu}, {Silva Aguirre},
  {Mazumdar}, {Mosumgaard}, {Lund}, \& {Ranadive}}]{verma++2019-helium}
{Verma}, K., {Raodeo}, K., {Basu}, S., {et~al.} 2019, \mnras, 483, 4678

\bibitem[{{Viani} {et~al.}(2017){Viani}, {Basu}, {Chaplin}, {Davies}, \&
  {Elsworth}}]{viani++2017-numax-sc-metal}
{Viani}, L.~S., {Basu}, S., {Chaplin}, W.~J., {Davies}, G.~R., \& {Elsworth},
  Y. 2017, \apj, 843, 11

\bibitem[{{Viani} {et~al.}(2018){Viani}, {Basu}, {Ong J.}, {Bonaca}, \&
  {Chaplin}}]{viani++2018-amlt-mh}
{Viani}, L.~S., {Basu}, S., {Ong J.}, M.~J., {Bonaca}, A., \& {Chaplin}, W.~J.
  2018, \apj, 858, 28

\bibitem[{{Virtanen} {et~al.}(2020){Virtanen}, {Gommers}, {Oliphant},
  {Haberland}, {Reddy}, {Cournapeau}, {Burovski}, {Peterson}, {Weckesser},
  {Bright}, {van der Walt}, {Brett}, {Wilson}, {Jarrod Millman}, {Mayorov},
  {Nelson}, {Jones}, {Kern}, {Larson}, {Carey}, {Polat}, {Feng}, {Moore}, {Vand
  erPlas}, {Laxalde}, {Perktold}, {Cimrman}, {Henriksen}, {Quintero}, {Harris},
  {Archibald}, {Ribeiro}, {Pedregosa}, {van Mulbregt}, \&
  {Contributors}}]{scipy}
{Virtanen}, P., {Gommers}, R., {Oliphant}, T.~E., {et~al.} 2020, Nature
  Methods, 17, 261

\bibitem[{{Vrard} {et~al.}(2016){Vrard}, {Mosser}, \&
  {Samadi}}]{vrard++2016-period-spacings}
{Vrard}, M., {Mosser}, B., \& {Samadi}, R. 2016, \aap, 588, A87

\bibitem[{{Wang} {et~al.}(2023{\natexlab{a}}){Wang}, {Huang}, {Zhou}, \&
  {Zhang}}]{wangc++2023-lamost-age}
{Wang}, C., {Huang}, Y., {Zhou}, Y., \& {Zhang}, H. 2023{\natexlab{a}}, \aap,
  675, A26

\bibitem[{{Wang} {et~al.}(2023{\natexlab{b}}){Wang}, {Li}, {Bi}, {Bedding}, \&
  {Li}}]{wangyx++2023-rgb}
{Wang}, Y., {Li}, T., {Bi}, S., {Bedding}, T.~R., \& {Li}, Y.
  2023{\natexlab{b}}, \apj, 953, 182

\bibitem[{{W}es {M}c{K}inney(2010)}]{pandas}
{W}es {M}c{K}inney. 2010, in {P}roceedings of the 9th {P}ython in {S}cience
  {C}onference, ed. {S}t\'efan van~der {W}alt \& {J}arrod {M}illman, 56 -- 61

\bibitem[{{Wu} {et~al.}(2018){Wu}, {Xiang}, {Bi}, {Liu}, {Yu}, {Hon}, {Sharma},
  {Li}, {Huang}, {Liu}, {Zhang}, {Li}, {Ge}, {Tian}, {Zhang}, \&
  {Zhang}}]{wuyq++2018-mass-rgb}
{Wu}, Y., {Xiang}, M., {Bi}, S., {et~al.} 2018, \mnras, 475, 3633

\bibitem[{{Xiang} \& {Rix}(2022)}]{xiang++2022}
{Xiang}, M., \& {Rix}, H.-W. 2022, \nat, 603, 599

\bibitem[{{Y{\i}ld{\i}z} {et~al.}(2016){Y{\i}ld{\i}z}, {{\c{C}}elik Orhan}, \&
  {Kayhan}}]{yildiz++2016-scaling-relation-tuning}
{Y{\i}ld{\i}z}, M., {{\c{C}}elik Orhan}, Z., \& {Kayhan}, C. 2016, \mnras, 462,
  1577

\bibitem[{{Ying} {et~al.}(2023){Ying}, {Chaboyer}, {Boudreaux}, {Slaughter},
  {Boylan-Kolchin}, \& {Weisz}}]{ying++2023-m92}
{Ying}, J.~M., {Chaboyer}, B., {Boudreaux}, E.~M., {et~al.} 2023, \aj, 166, 18

\bibitem[{{Yu} {et~al.}(2020){Yu}, {Bedding}, {Stello}, {Huber}, {Compton},
  {Gizon}, \& {Hekker}}]{yuj++2020-lpv1}
{Yu}, J., {Bedding}, T.~R., {Stello}, D., {et~al.} 2020, \mnras, 493, 1388

\bibitem[{{Yu} {et~al.}(2018){Yu}, {Huber}, {Bedding}, {Stello}, {Hon},
  {Murphy}, \& {Khanna}}]{yuj++2018-16000-rg}
{Yu}, J., {Huber}, D., {Bedding}, T.~R., {et~al.} 2018, \apjs, 236, 42

\bibitem[{{Zhou} {et~al.}(2020){Zhou}, {Asplund}, {Collet}, \&
  {Joyce}}]{zhouyx++2020-rg}
{Zhou}, Y., {Asplund}, M., {Collet}, R., \& {Joyce}, M. 2020, \mnras, 495, 4904

\bibitem[{{Zhou} {et~al.}(2023){Zhou}, {Christensen-Dalsgaard}, {Asplund},
  {Li}, {Trampedach}, {Ting}, \& {R{\o}rsted}}]{zhouyx++2023-numax}
{Zhou}, Y., {Christensen-Dalsgaard}, J., {Asplund}, M., {et~al.} 2023, arXiv
  e-prints, arXiv:2310.20050

\bibitem[{{Zinn} {et~al.}(2023){Zinn}, {Pinsonneault}, {Bildsten}, \&
  {Stello}}]{zinn++2023-adiabatic}
{Zinn}, J.~C., {Pinsonneault}, M.~H., {Bildsten}, L., \& {Stello}, D. 2023,
  \mnras, 525, 5540

\bibitem[{{Zinn} {et~al.}(2019){Zinn}, {Pinsonneault}, {Huber}, {Stello},
  {Stassun}, \& {Serenelli}}]{zinn++2019-radius-sc}
{Zinn}, J.~C., {Pinsonneault}, M.~H., {Huber}, D., {et~al.} 2019, \apj, 885,
  166

\end{thebibliography}
\bibliographystyle{apj}



\end{document}